\documentclass[fleqn,usenatbib,useAMS]{mnras}
\usepackage{amssymb}
\usepackage{graphics,epsfig,psfig}
\usepackage{dblfloatfix}
\usepackage{graphicx}
\usepackage[normalem]{ulem}
\usepackage[dvipsnames]{color}
\usepackage{xcolor,subfig}
\usepackage{amsmath, amssymb}
\usepackage[]{inputenc,amssymb}
\usepackage{multirow, multicol}
\usepackage{booktabs,caption}
\usepackage{lipsum}
\usepackage{gensymb}
\usepackage[export]{adjustbox}
\usepackage{mathtools} 
\usepackage{makecell}

\definecolor{ao(english)}{rgb}{0.0, 0.5, 0.0}
\definecolor{cadmiumgreen}{rgb}{0.0, 0.42, 0.24}
\definecolor{awesome}{rgb}{1.0, 0.13, 0.32}
\definecolor{dogwoodrose}{rgb}{0.84, 0.09, 0.41}
\definecolor{electricgreen}{rgb}{0.0, 1.0, 0.0}
\definecolor{forestgreen(web)}{rgb}{0.13, 0.55, 0.13}

\def \be{\begin{equation}}
\def \ee{\end{equation}}
\def \ba{\begin{eqnarray}}
\def \ea{\end{eqnarray}}

\def \sfr{M_\odot$ yr$^{-1}}

\newcommand{\divr}{ {\nabla}\cdot}

\title[CR driven galactic outflow]{Role of cosmic rays in the early stages of galactic outflows}
\author[Jana, Gupta, Nath]
{Ranita Jana$^1$\thanks{Email: ranita@rri.res.in}, Siddhartha Gupta$^{1,2,3}$, Biman B. Nath$^{1}$\\
$^1$ Raman Research Institute, Sadashiva Nagar, Bangalore 560080, India\\
$^2$ Department of Physics, Indian Institute of Science, Bangalore, India\\
$^3$ Department of Astronomy \& Astrophysics, University of Chicago, IL 60637, USA\\
}
\begin{document}
\maketitle
\label{firstpage}
\begin{abstract}
Using an idealized set-up, we investigate the dynamical role of cosmic rays (CRs) in the early stages of galactic outflows for galaxies of halo masses $10^{8}$, $10^{11}$ and $10^{12}$ $M_\odot$. The outflow is launched from a central region in the galactic disk where we consider three different constant star formation rates ($0.1$, $1$, and $10$ $\sfr$) over a dynamical timescale of $50$ Myr. We determine the temperature distribution of the gas and find that CRs can reduce the temperature of the shocked gas, which is consistent with previous results. However, we show that CRs do not have any noticeable effect on the mass loading by the outflow. We find that CRs can {\it reduce} the size of the outflow, which contradicts previous claims of efficient dynamical impact of CRs; however, it is consistent with earlier theoretical models of cosmic ray driven blastwave as well as stellar wind. We discuss the dependence of our results on  CR injection prescriptions and compare them with earlier studies. We conclude that in the early stages of galactic outflows the dynamical role of CRs is not important.
\end{abstract}
\section{Introduction}\label{sec:intro}
Galaxies form as a consequence of gravitational collapse of large gas clouds into dark matter halos. With the onset of star formation, a substantial amount of gas is expelled from the galaxies as the form of biconical outflow, as attested by observations of starburst galaxies (see e.g., \citealt{Veilleux2005} for review). These galactic scale outflows play an important role in the dynamical and chemical evolution of galaxies. Theoretical studies that do not include such feedback mechanism overproduce the star formation rate (SFR) and baryon fraction in galaxies. Outflows are therefore believed to regulate the star formation process, and consequently influence galactic evolution. These outflows also enrich the inter-galactic matter with metals which mostly form in the star forming regions. Although outflows are a crucial aspect of galactic evolution, their driving mechanism and modes of propagation are still poorly understood.

Thermal pressure due to shock heating of the inter-stellar medium (ISM) by supernovae and stellar winds in star formation sites have been the focus of most previous studies (\citealt{Efstathiou2000,Girichidis2016a,Martizzi2016}). Non-thermal pressure, such as pressure due to radiation from young stars (\citealt{Murray2005,Hopkins2012, Krumholz2012, Agertz2013,Skinner2015,Rosdahl2015}) and/or cosmic rays (CRs) (\citealt{Ipavich1975,Drury1981,Breitschwerdt1991,Zirakashvili1996,Hanasz2003,Samui2010,Recchia2016,Gupta2018Jan}) have also been suggested as important driving mechanism as thermal pressure. Preliminary analytical works assumed ideal coupling between CRs and thermal gas mediated by damped Alfv\'en waves and without diffusion. \citet{Ipavich1975} considered a spherical outflow emanating from a point mass galaxy of $10^{11}$ $M_\odot$ and found that CRs escaping from a galaxy are able to carry thermal gas with them and produce a galactic wind with mass loss rate $1\hbox{--}10$ $\sfr$. \citet{Breitschwerdt1991} improved this model by considering a disk galaxy and a `mushroom'-type geometry of the outflow. \citet{Everett2008} 
used these ideas for a possible outflow from Milky Way and reported that galactic wind models incorporating thermal pressure as well as CR pressure produce the best fit to the observed Galactic diffuse soft X-ray emission. Recently, \citet{Samui2018} used a spherically symmetric thin shell model to study the effect of CR pressure on the dynamics of outflow. They suggested that for the low mass galaxies where thermal pressure alone cannot sustain a large scale galactic outflow (because of radiative cooling, which peaks at $\sim 10^5$ K), the inclusion of CRs may drive a steady outflow.

In addition to analytical models, hydrodynamical (HD) and magnetohydrodynamical (MHD) simulations have also been used. \citet{Uhlig2012} performed the first HD simulations to investigate the effects of CR heating due to streaming instability (see eg. \citet{Kulsrud1969} for review). However, CR diffusion was not included in their simulations and CRs were assumed to heat the gas above the disk scale height, which can affect the outflow. 
For a better understanding, \citet{Booth2013} and \citet{Salem2014} used a more realistic set-up by considering advection as well as isotropic diffusion of cosmic rays seft-consistently. They found that the large-scale pressure gradient established by CR diffusion helps to drive the wind. \citet{Ruszkowski2017} performed global three dimensional MHD simulation of an isolated Milky Way sized galaxy  to investigate the role of two different CR transport mechanisms, namely streaming and anisotropic diffusion. Similar to \citet{Booth2013} they found that SFR is significantly decreased when CRs are included in their simulations. They also found that in the presence of moderately super-Alfvenic CR streaming, the mass loading factor ranges between $ \sim 0.25$ to $\sim 0.6$. Recently, these models have been further improved by \citet{Butsky2018} where CR streaming, isotropic and anisotropic diffusion of cosmic rays have been investigated. They reported that all three transport mechanisms result in strong metal rich outflows which largely differ from the models without CRs in terms of the temperature and ionization structure of circumgalactic medium (CGM). 
 
The effects of CRs are primarily quantified using the parameters such as mass loading factor and the presence of multiphase gas in the outflow and in the CGM. The mass loading factor (usually denoted by $\eta$), which is defined as the ratio of gas outflow rate to the SFR, and is indicative of the efficiency of outflows in ejecting mass from the host galaxy. For example, \citet{Booth2013} found that $\eta \sim 0.5$ in the case of Milky Way-sized galaxies (virial mass of $10^{12}$ $M_\odot$), with or without CRs, while in the case of low mass galaxies, such as the Small Magellanic Cloud (SMC) ($2 \times 10^9$ $M_\odot$), it can be $\sim 10$ with CR and $\sim 1$ without CRs. A multiphase structure of galactic winds and the existence of cold gas ($T \sim 10^4 K$) in the wind was noticed by \citet{Booth2013}, \citet{Salem2014} and \citet{Butsky2018}. This owes to the fact that while thermal pressure driven winds are accelerated in the disk, the CR driven winds are accelerated smoothly into the halo. The pressure gradient in the halo is  $3\hbox{--}10$ times higher in the case of CR driven wind, compared to pure thermal pressure driven wind. They have shown that the CR-driven winds have a lower speed and support `cold' gas, which the authors have used to explain the existence of multiphase gas in the CGM.  
CR driven winds in massive galaxies have been recently studied by \citet{Fujita2018}, for a galaxy of mass $5 \times 10^{12}$ $M_\odot$, with a continuous mechanical power of $10^{43}$ erg s$^{-1}$, that corresponds to a SFR of $\sim 100\hbox{--}500$ $\sfr$. They found it is difficult for CRs to drive a wind from massive galaxies, even with a large SFR, and found that $\eta \lesssim 0.006\hbox{--}0.03$. In addition to this, \citet{Jacob2018} performed simulations of galaxies of mass range between $10^{10}$ and $10^{13}$ $M_\odot$ and concluded that mass loading factor drops rapidly with virial mass with an approximate relation $\eta \sim$ $M^{\alpha}_{\rm vir}$ where $\alpha$ is between $-1$ and $-2$. This relation is slightly steeper than the previously reported results. Their simulations also reveal that CRs cannot drive a steady mass-loaded outflow if the virial mass of the galaxy is more than $10^{12}$ $M_\odot$ which supports the results by \citet{Fujita2018}.
 
Most of the previous works, with the notable exception of \citet{Fujita2018}, assume a feedback mechanism by which the SFR is regulated, and the star formation activity changes with time during the simulation runs. The resulting effects of CR is therefore entangled with various factors and it is not clear how (a) galactic mass, (b) star formation rate, (c) CR injection sites and (d) different assumptions of diffusion contribute towards the propagation of outflows. It is no wonder that the results of these studies  have remained inconclusive. 

In this paper, we study with the help of idealized simulations, the outflow properties for three galactic masses ($10^{12}$, $10^{11}$ and $10^{8}$ $M_\odot$, hereafter M12, M11 and M8 respectively) using constant star formation rates for $\sim$ few Myr. Our primary focus is to investigate the outflow dynamics in the early stages (until $50$ Myr) of galaxy evolution, where the constant star formation rate is a reasonable assumption. For the fiducial case, we also investigate long term evolution (until $210$ Myr) with time-dependent star formation rate, however, without including feedback from the outflowing gas on the star formation. These assumptions allow us to distinguish the role of the individual physical process separately, which can be extended in a more realistic scenario.
 
We find that although the presence of CRs increases the cold gas mass in the outflow, the dynamical impact of CRs is not important in the early stages of galaxy evolution. 
We present our simulation set up and results in Sections \ref{sec:setup} and \ref{sec:results}. We discuss the results of our long-duration simulations with periodic star formation in Section \ref{sec:long_dur}. Implications of our work and comparisons with previous studies are discussed in Section \ref{sec:discussion} and summarised in Section \ref{sec:summary}.
 
\section{Simulation set-up}\label{sec:setup}
We solve the following two-fluid CR hydrodynamical equations using \textsc{pluto} code (\citealt{Mignone2007, Gupta2019})
\begin{flalign}
\label{eq:masscont}
\frac{\partial \rho}{\partial t} + \nabla\cdot (\rho \boldsymbol{v}) = S_{\rho} &&\\
\label{eq:momcont}
\frac{\partial (\rho {\boldsymbol v})}{\partial t} + \divr ({\rho {\boldsymbol v} \otimes {\boldsymbol v}}+ p_{\rm t} {\bf I} )  + \rho \nabla \Phi_{\rm t} - \frac{\rho v_{\phi}^2}{R} \hat{\bf R} = 0 && \\
\label{eq:energytot}
\frac{\partial }{\partial t}(\rho v^2/2 +e_{\rm th} + e_{\rm cr} ) + \divr [(\rho v^2/2+ & & \nonumber\\ e_{\rm th} + e_{\rm cr}   + p_{\rm t}) {\boldsymbol v}]   =  -q_{\rm cool}  + S_{\rm th} &&\\
\label{eq:energycr}
\frac{\partial e_{\rm cr}}{\partial t} + \divr({e_{\rm cr} {\boldsymbol v}}+{\bf F}_{\rm crdiff})  =  - p_{\rm cr} \divr {\boldsymbol  {v}} +S_{\rm cr} &&
\end{flalign}
Here $\rho$ is the mass density and $v$ is the velocity of the gas. The total pressure, $p_{\rm t}$, is the sum of thermal pressure ($p_{\rm th}$) and CR pressure ($p_{\rm cr}$). 
Here $\rho v^2/2$ is the kinetic energy density, $e_{\rm th} = p_{\rm th}/ (\gamma_{\rm th}-1)$ is the thermal energy density and  $e_{\rm cr} = p_{\rm cr}/ (\gamma_{\rm cr} -1)$ is the CR energy density , where $\gamma_{\rm th} = 5/3$ and $\gamma_{\rm cr} = 4/3$ are the adiabatic constants of the gas and CRs respectively. The term $v_{\phi}^2/R\hat{\bf R}$ in the momentum equation represents the centrifugal force that acts on the gas due to rotation of the disk.
 $\Phi_{\rm t}$ is the total gravitational potential (cf. Section \ref{sec:gravpoten}). $S_{\rm \rho}$, $S_{\rm th}$ and $S_{\rm cr}$ are the injected mass and energy source terms. The term, $q_{\rm cool}$, denotes the energy lost due to radiative cooling of thermal gas. $\bf {F}_{\rm crdiff}$ represents the flux term associated with isotropic CR diffusion.

\subsection{Gravitational Potential}\label{sec:gravpoten}
The total gravitational potential denoted by $\Phi_{\rm t}$ in equations \ref{eq:momcont} is the superposition of gravitational potential exerted by a stellar disk and dark matter halo. 
For the stellar disk, we use the Miyamoto \& Nagai potential \citep{Miyamoto1975}. In cylindrical coordinate ($R,z$),  it is given by
\begin{equation}
\Phi_{\rm disk}(R,z) = - \frac{G M_{\rm disk}^\star}{\sqrt{R^2 + (a + \sqrt{z^2 +b^2})^2)}} , \hspace{0.3cm}(a,b \geq 0),
\end{equation}
Here $R = r \sin \theta$ and $z = r \cos\theta$. $M_{\rm disk}^\star$ is the mass of the stellar disk. The two parameters $a$ and $b$ represent the scale length and scale height of the disk respectively. 

For the dark matter (DM) halo, we use a modified Navarro-Frenk-White Model \citep{Navarro1997}. The potential is given as 
\begin{equation}
\Phi_{\rm DM} = -\frac{G M_{\rm vir}}{\sqrt{R^2+z^2+d^2}}\Bigg[\frac{\ln(1+\sqrt{R^2+z^2+d^2}/r_s)}{f(c)}\Bigg]\hspace{0.3 cm} (d \geq 0),
\end{equation}
where $f(c)=\ln(1+c) - {c \over (1+c)}$ with $c=r_{\rm vir}/r_s$, the concentration parameter. $M_{\rm vir}$ is the total mass of the galaxy (including DM) within virial radius ($r_{\rm vir}$). $r_s$ and $d$ are the scale radius and core radius of the DM distribution respectively. These parameters are listed in Table \ref{tab:modelpara}. 

\subsection{Initial density and pressure distribution}\label{sec:ini_dist}
The initial gas distribution in the galaxy is assumed to have two components: (i) a warm ionized gas (few times $10^4$ K) in the disk and (ii) a hot gas ($\sim T_{\rm vir}$). In order to set the initial gas distribution, we use the steady state solution of the Euler's equation by solving 
\begin{equation}
- \frac{\nabla p}{\rho} - \nabla \Phi_{\rm t} + \frac{v_{\phi}^2}{R} \hat{R} = 0  \ ,
\end{equation}
for each component. The disk and halo gas mass distributions are found to be
\begin{eqnarray}\label{eq:rho_d}
\rho_{\rm d}(R,z) & = & \rho_{\rm d0} \exp\big[-\frac{1}{{a_{\rm sd}^2}}\big\{\Phi_{\rm t}(R,z)- \Phi_{\rm t}(0,0) - \nonumber\\
 & & \hspace{1.4cm} f^2(\Phi_{\rm t}(R,0)-\Phi_{\rm t}(0,0))\big\}\big]
\end{eqnarray}
 and 
\begin{eqnarray}\label{eq:rho_h}
\rho_{\rm h}(R,z) = & \rho_{\rm h0} \exp\Big[- \frac{1}{a_{\rm sh}^2} \big\{\Phi_{\rm t}(R,z)- \Phi_{\rm t}(0,0)\big\}\big]
\end{eqnarray}
respectively. Here $\rho_{\rm d0}$ and $\rho_{\rm h0}$ are the central density, and $a_{\rm sd}$ and $a_{\rm sh}$ are isothermal sound speed of the warm disk gas and hot halo gas respectively. The two components of the thermal gas (i.e., the warm and hot ionized gas) in the simulation have been treated as a single fluid. Therefore, the total density of the thermal gas is given by $\rho = \rho_{\rm d} + \rho_{\rm h}$. 

The term, $f$ in Eq. \ref{eq:rho_d} determines the centrifugal force on the gas due to the rotation of stellar disk, i.e., 
\begin{equation}
v_{\phi} = f \Big[\frac{\rho_d}{\rho} {\rm R} \Big(\frac{\partial \Phi_{\rm t}}{\partial R}\Big)_{z=0}\Big]^{1/2}
\end{equation}

In the thermally driven outflow models, initially, total pressure of the gas is the sum of pressure due to the disk gas and the halo gas. In the models, where CRs are included, assuming equipartition of energy in the disk, we divide the total thermal pressure in the disk such that $p_{\rm th} = 2\ p_{\rm cr}$. We assume that initially halo gas has a very low CR pressure. 

\begin{figure*}
\includegraphics[width=2.35 in, height = 1.5 in]{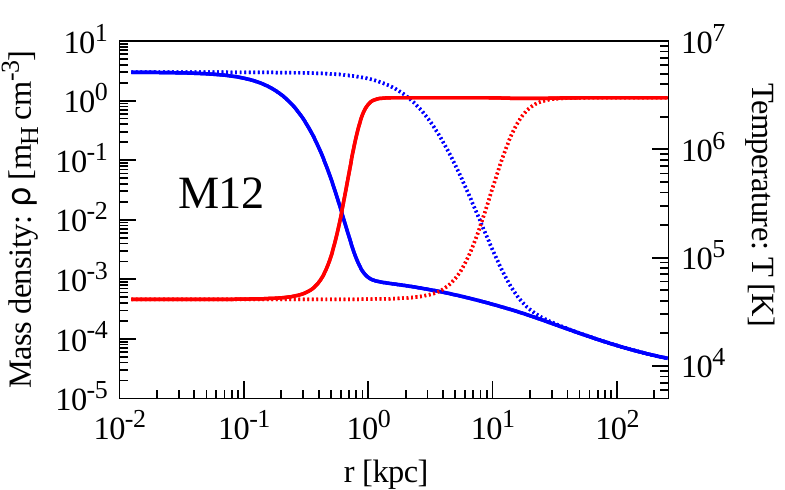}%
\includegraphics[width= 2.35 in, height = 1.5 in]{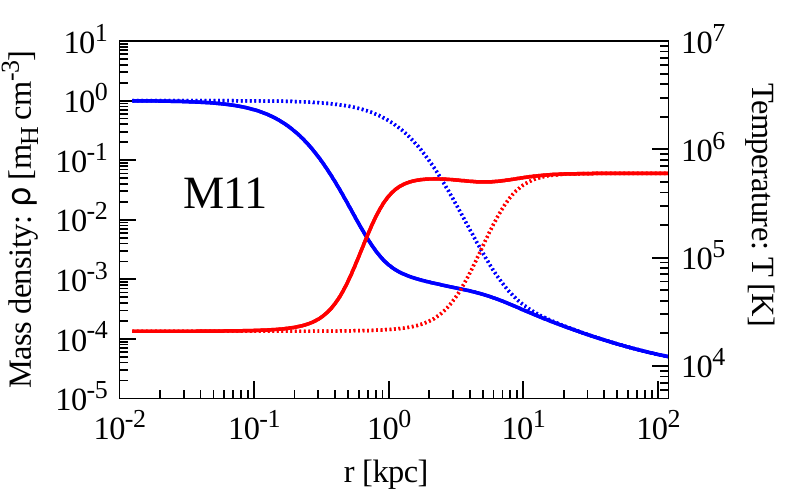}%
\includegraphics[width= 2.35 in, height = 1.5 in]{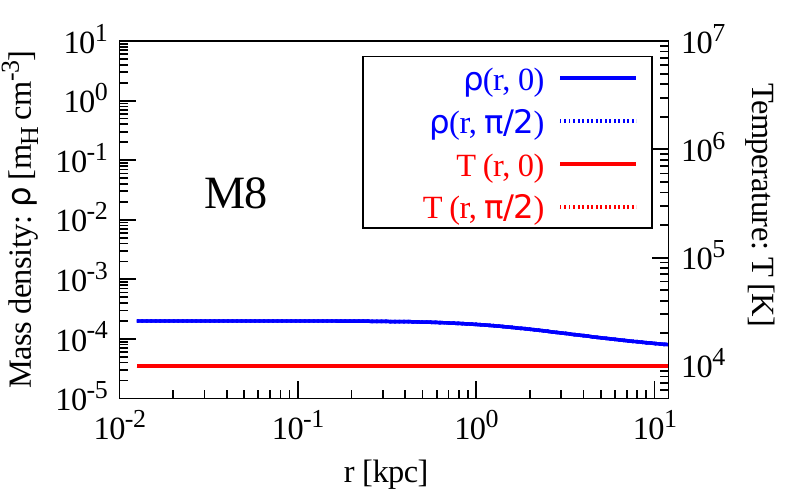}%
\caption{Initial density and temperature distributions for three galaxies with halo mass (from left) $10^{12} M_\odot$, $10^{11} M_\odot$ and $10^{8} M_\odot$ respectively. Density (blue) and temperature (red) profiles are shown along $\theta=0$ and $\theta=\pi/2$. For $M_{\rm vir}=10^{8}\,  M_\odot$ (rightmost panel), the profiles are isotropic, because any disk in such a galaxy is expected to be very small compared to the outflow length scale of interest and have not been considered in our simulation.}
\label{ini_den}
\end{figure*}

The set-up discussed above is similar to \citep{Sarkar2015}, except that here we have CRs as an additional fluid. We find that for all model parameters listed in Table \ref{tab:modelpara}, the gas distribution is reasonably stable until $t \gg 100$ Myr. The timescale we are interested in is much smaller than this. The initial density and temperature profiles of the three cases are shown in Figure \ref{ini_den}.

\begin{center}
\begin{table}
\caption{The values of parameters used in our simulations}
\label{tab:modelpara}
\begin{tabular}{cccc}
\hline
\hline
Parameters & Values & Values & Values\\
(Units) &  (M12) & (M11) & (M8)\\
\hline
\\
$M_{\rm vir}\,(M_\odot)$ & $10^{12}$ & $10^{11}$& $10^{8}$\\
$M_{\rm disk}^\star\,(M_\odot)$ & $5 \times 10^{10}$ & $5 \times 10^{9}$& $--$\\
$T_{\rm vir}\,(\rm K)$ & $3 \times 10^6$ & $6 \times 10^5$& $10^{4}$\\
$T_{\rm disk}\,(\rm K)$ & $4 \times 10^4$ & $2 \times 10^4$& $--$\\
$r_{\rm vir}\,(\rm kpc)$ & $258$ & $120$& $12$\\
$r_{\rm s}\,(\rm kpc)$ & $21.5$ & $10.0$ & $1.0$\\
$a\,(\rm kpc)$ & $4.0$ & $2.0$& $--$\\
$b\,(\rm kpc)$ &$ 0.4$ & $0.2$& $--$\\
$d\,(\rm kpc)$ & $6.0$ & $2.0$& $1.0$\\
$c$ & $12$ & $12$ & $12$\\
$f$ & $0.95$ & $0.9$ & $--$\\
$Z_{\rm disk}\,(\rm Z_\odot)$ & $1.0$ & $1.0$& $--$\\
$Z_{\rm halo}\,(\rm Z_\odot)$ & $0.1$ & $0.1$& $0.1$\\
$\rho_{\rm d0}\,(m_{\rm H} \rm cm^{-3})$ & $3.0$ & $1.0$& $--$\\
$\rho_{\rm h0}\,(m_{\rm H} \rm cm^{-3})$ & $1.1 \times 10^{-3}$ & $1.5 \times 10^{-3}$ & $2 \times 10^{-4}$\\
\\
\hline
\hline
\end{tabular}
\end{table}
\end{center}

\subsection{Solver, grid distribution and boundary}\label{sec:grid}
We perform our simulations in 2D spherical geometry with three velocity components. To solve the two-fluid hydrodynamical equations, we use the HLL Riemann solver and take piecewise linear spatial reconstruction for all variables. Time has been evolved using  Runge-Kutta 2nd order scheme. We define the computational box from $0.01$ kpc to $30$ kpc in the $r$ direction and from $0.01$ to $\pi/2$ in the $\theta$ direction. In the $r$ direction, we take a uniform grid up to $110$ pc (20 grid points) and a logarithmic grid thereafter such that total grid along $r$ is $512$. Along the $\theta$ direction, we take $256$ uniform grid distribution. For the extended run, we choose a larger box (70 kpc) with 1024 grid points in the $r$ direction. The inner and outer boundaries for $r$ are set to be outflow, and for $\theta$, both inner/outer boundaries are reflective.

\subsection{Cooling}\label{sec:cooling}
We have considered radiative cooling of the warm disk gas and hot halo gas using the tabulated cooling function from \citet{Sutherland1993}. The dependence of cooling functions on metallicity is taken into account using linear interpolation of the cooling curves between two metallicities. We have assumed that the metallicity of disk gas is solar ($Z = Z_\odot$) and the halo gas metallicity is $Z = 0.1\,Z_\odot$.

In the simulation, the disk gas rapidly loses its thermal energy due to high metallicity and high density when cooling is considered. However, in reality the disk gas is maintained at $\sim 10^4$ K by stellar radiation. Since we do not include radiation in our simulation, we use a box size of $\rm R \times z = 15 \times 2$ $\rm kpc^2$ for the $10^{12}$ $M_\odot$ galaxy and a box size of $\rm R \times z = 10 \times 4 $ $\rm kpc^2$ for the $10^{11}$ $M_\odot$ galaxy within which we do not allow the disk gas (initially kept at a few times $10^4$ K) to cool. However, we allow the injected material to cool, but not below a radiative cooling floor of $10^4$ K. (see Appendix \ref{app:coolbox} for details of this cooling constraint.)

\subsection{Mass and energy injections}\label{sec:injection}
We assume that multiple supernovae explosions in the central region of the galaxy give rise to continuous energy input in the form of a constant mechanical luminosity. The central region  within which thermal energy is deposited is fixed by the condition proposed by \cite{Sharma2014} so that the cooling rate is smaller than the energy deposition rate. The injection region is of radius, $r_{\rm inj} = 30$ pc in all our models. We perform simulations with three different star formation rates for each galaxy. For Salpeter IMF, assuming the lower and upper limits of stellar mass as 0.1 and 100 $M_\odot$ and because each supernova gives rise to $\sim 10^{51}$ erg energy, the relation between mechanical luminosity and SFR (\citealt{Salpeter1955, Strickland2007}) is given by
\begin{equation}
L = 7 \times 10^{40} \mathrm{erg \, s^{-1}} \left(\frac{\rm SFR}{1\, M_\odot \rm yr^{-1}}\right)
\end{equation}
On an average $10\%$ of the SFR is the rate of mass injection into the interstellar matter i.e. $\dot{M}_{\rm inj} = 0.1 \times \rm SFR$.

CRs are injected in the form of pressure wherever a shock is detected (see Fig. \ref{fig:M12_compare}) since shocks are the acceleration sites for CRs. The following conditions \citep{Gupta2018Oct} have been used to detect whether a particular region within the simulation box is shocked or not
\begin{enumerate}
\item $\divr \boldsymbol{v} < 0$
\item $\Delta x |\nabla p| / p > \delta_{\rm tolerance}$
\item $\nabla T . \nabla \rho > 0$
\end{enumerate}
For all our simulations we have used $\delta_{\rm tolerance} = 0.5$. Wherever a shock is detected we redistribute the total pressure between thermal and CR component following the CR pressure component fraction $w = p_{\rm cr}/(p_{\rm th} + p_{\rm cr})$. In all our models we have used $w = 0.2$.

\section{Results}\label{sec:results}
To understand the  properties of outflowing gas between the models with and without CRs, we investigate (i) the morphology of the outflowing gas, (ii) outer shock position, (iii) mass-loading factor, and (iv) temperature distribution of the gas. These quantities have been used in previous studies to quantify the dynamical impact of CRs. Here we present the results from our simulations of three different galaxies with halo masses $M_{\rm vir}=10^{8}\, , 10^{11}$ and $10^{12}\, M_{\odot}$ using three different constant star formation rates (SFRs)\footnote{It is to be noted that for $10^{8}$ $M_\odot$ galaxy, SFRs $0.1$ and $1$ $\sfr$ are considered since $10$ $\sfr$ is  unlikely in such a small galaxy.}: $0.1, 1$ and $10$ $M_\odot \, \rm yr^{-1}$. To represent runs with and without CRs, we have used two different labels: `TH+CR' and `TH' respectively. Note that we have used our simulation for M11 galaxy as the fiducial run. The used parameters are given in Table \ref{tab:modelpara}. 

\subsection{Outflow morphology}\label{sec:morphology}
\begin{figure*}
\centering
\includegraphics[width=2.6in,height= 4.2in]{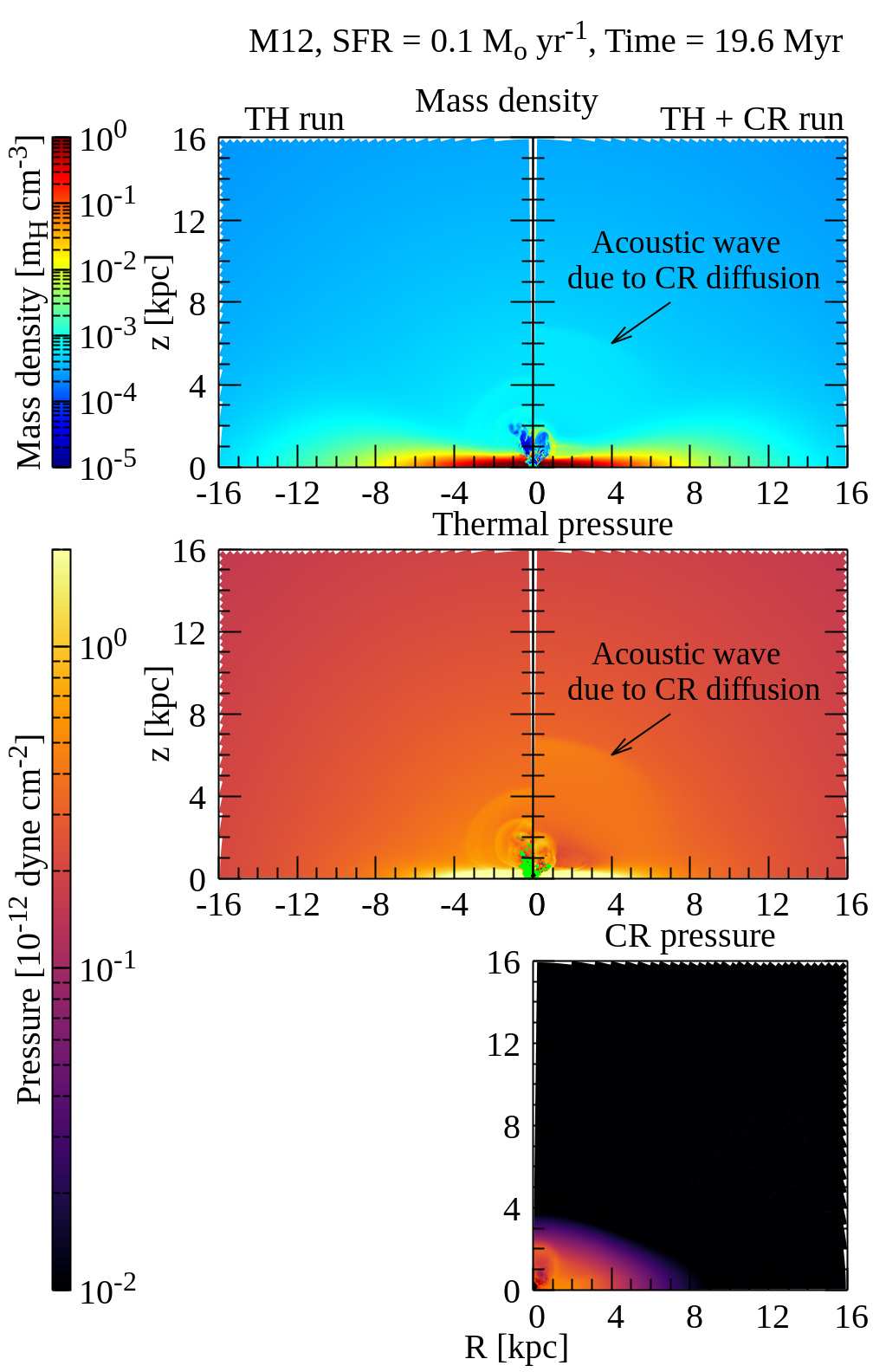}%
\includegraphics[width=2.2in,height= 4.2in]{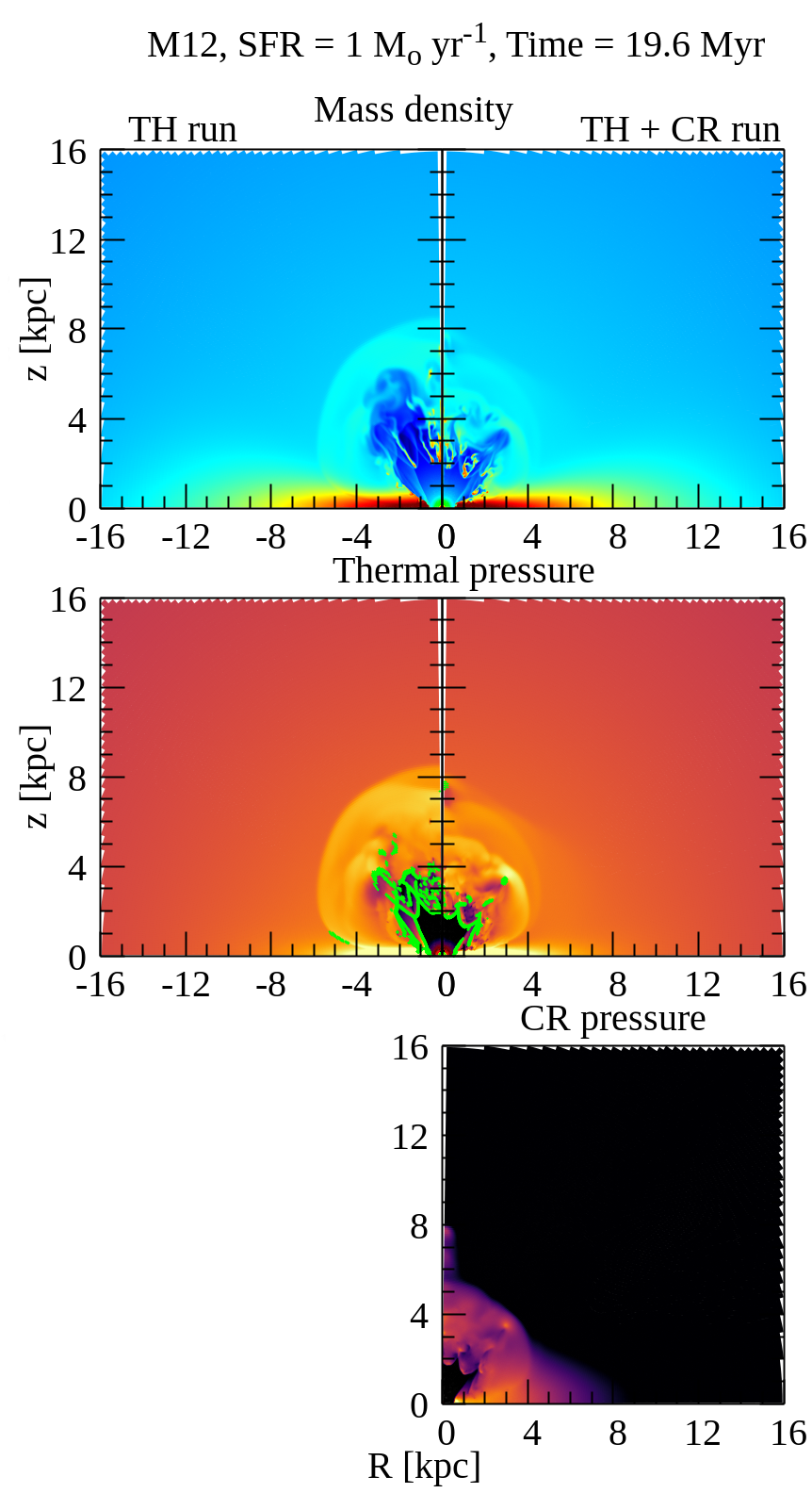}%
\includegraphics[width=2.2in,height= 4.2in]{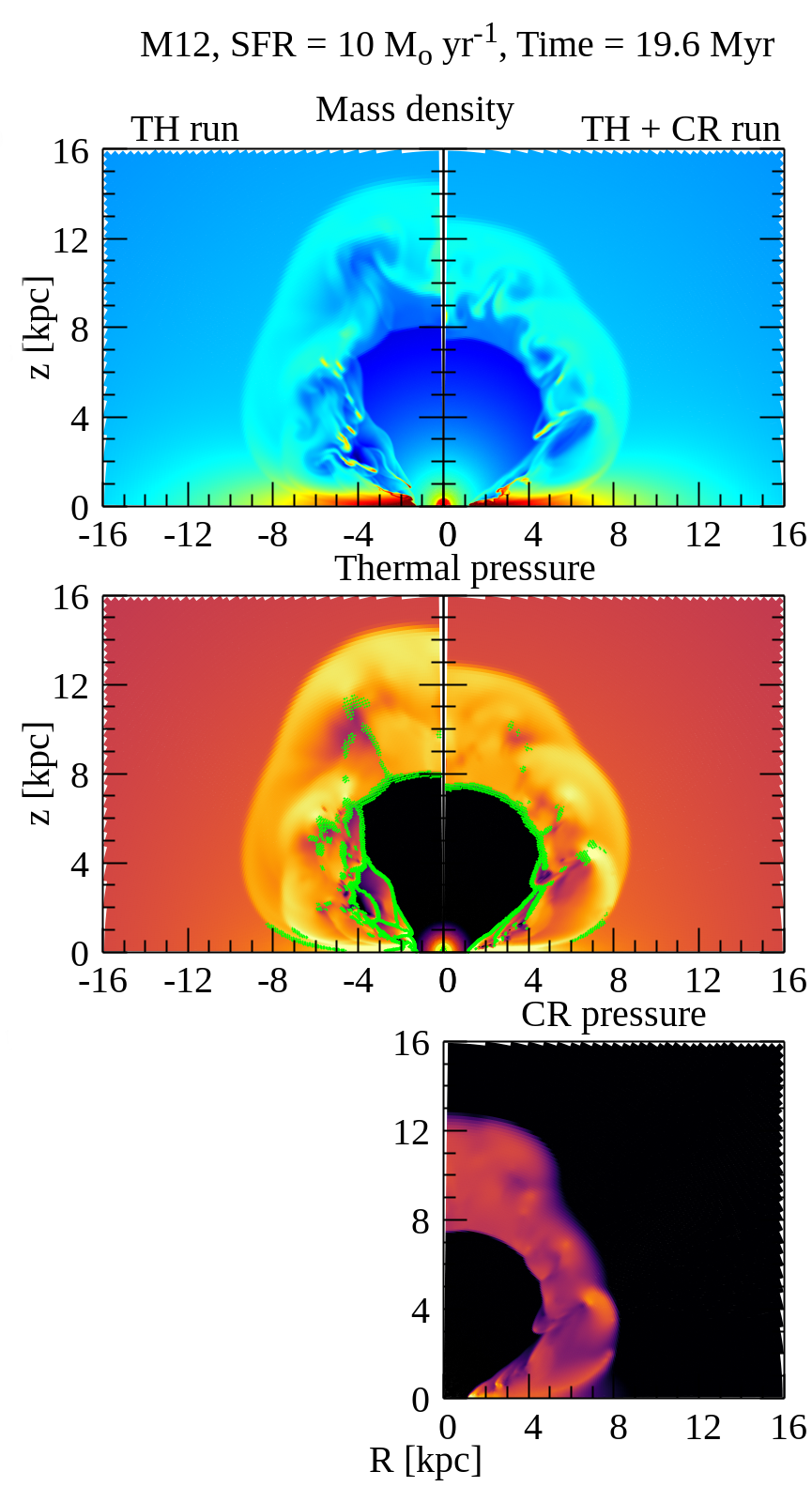}%
\caption{Comparison of density and thermal pressure between TH and TH+CR runs for M12 galaxy at $\approx 20$ Myr simulation time. The three set of plots (from left) are for SFR = 0.1, 1 and 10 $\sfr$. We also show the CR pressure in a separate plot. Green dots show the shock locations. In the TH+CR runs, these are the locations where CR energy is injected.}
\label{fig:M12_compare}
\centering
\includegraphics[width=2.6in,height= 4.2in]{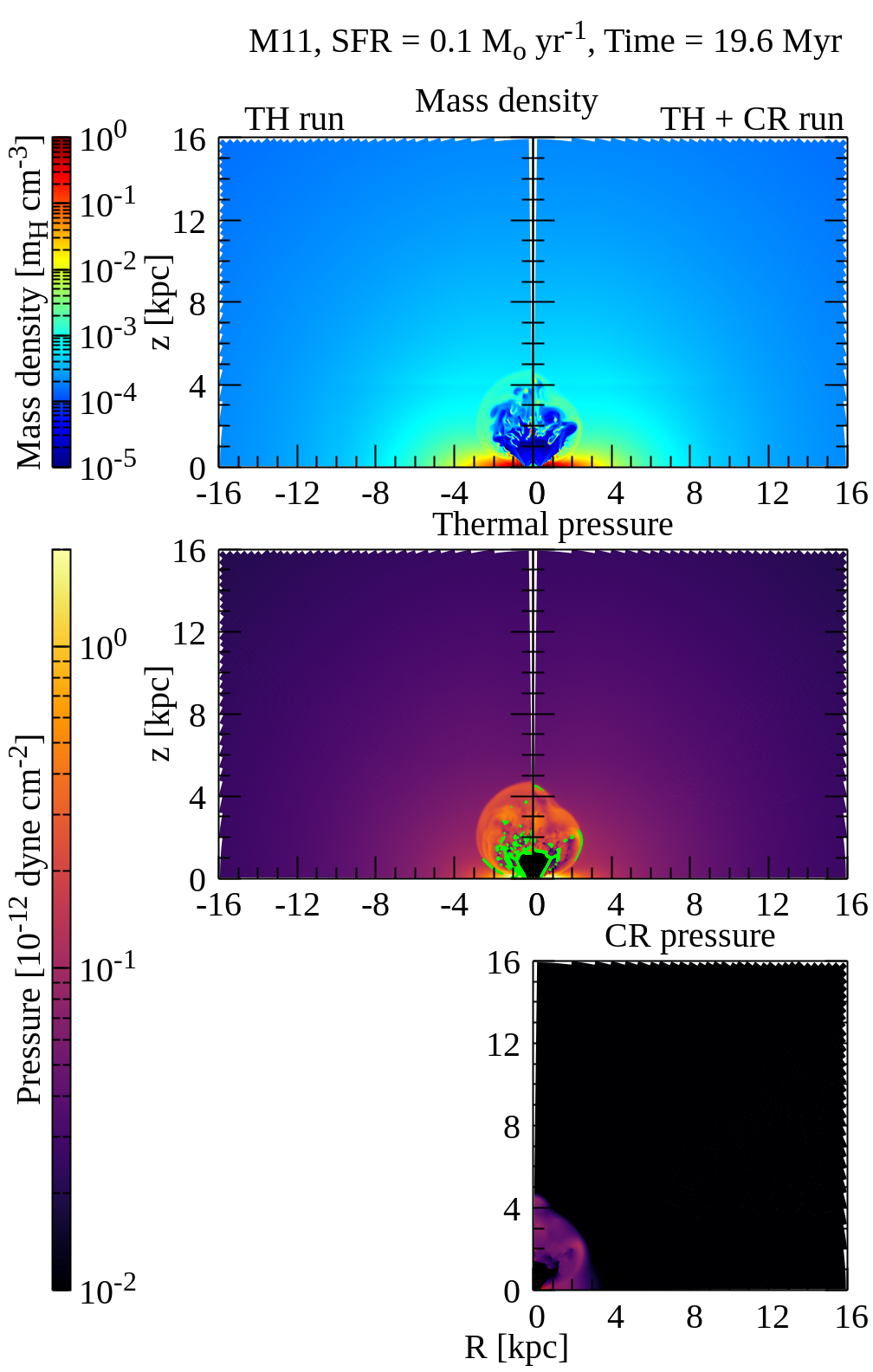}%
\includegraphics[width=2.2in,height= 4.2in]{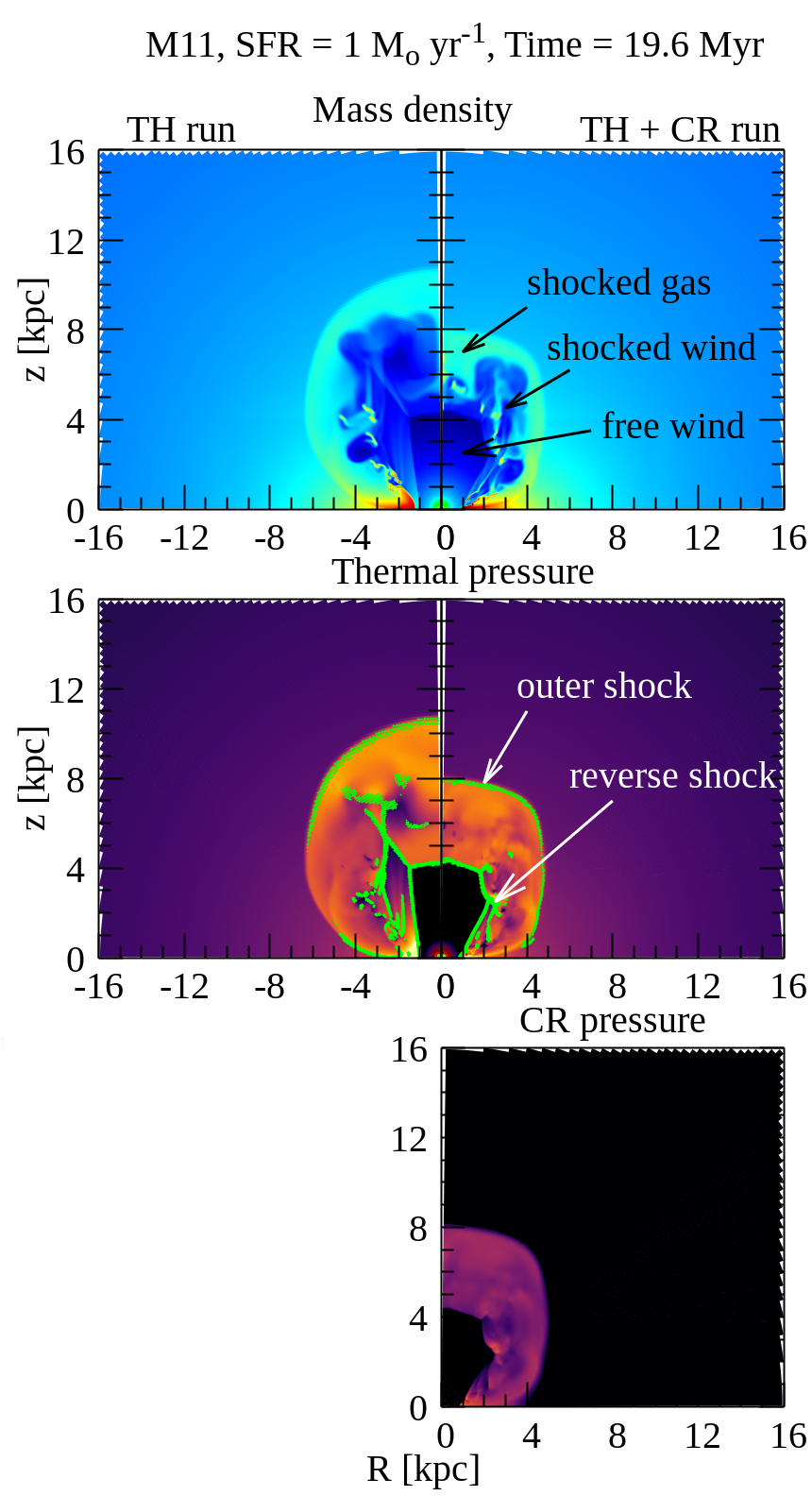}%
\includegraphics[width=2.2in,height= 4.2in]{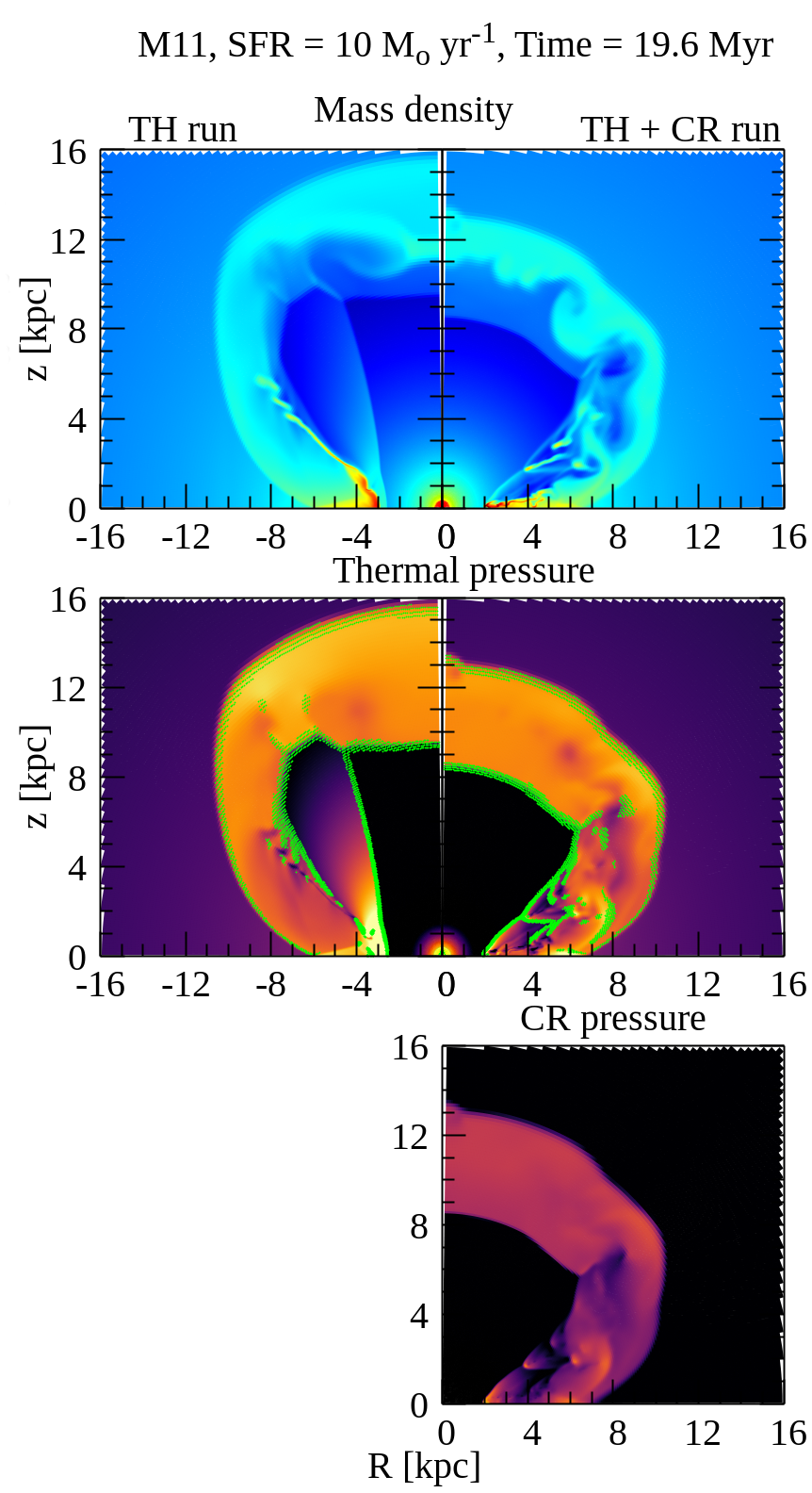}%
\caption{Same as Fig. \ref{fig:M12_compare}, for $M_{\rm vir} = 10^{11} M_\odot$} 
\label{fig:M11_compare}
\end{figure*}

\begin{figure*}
\includegraphics[width=2.6in,height= 4.in]{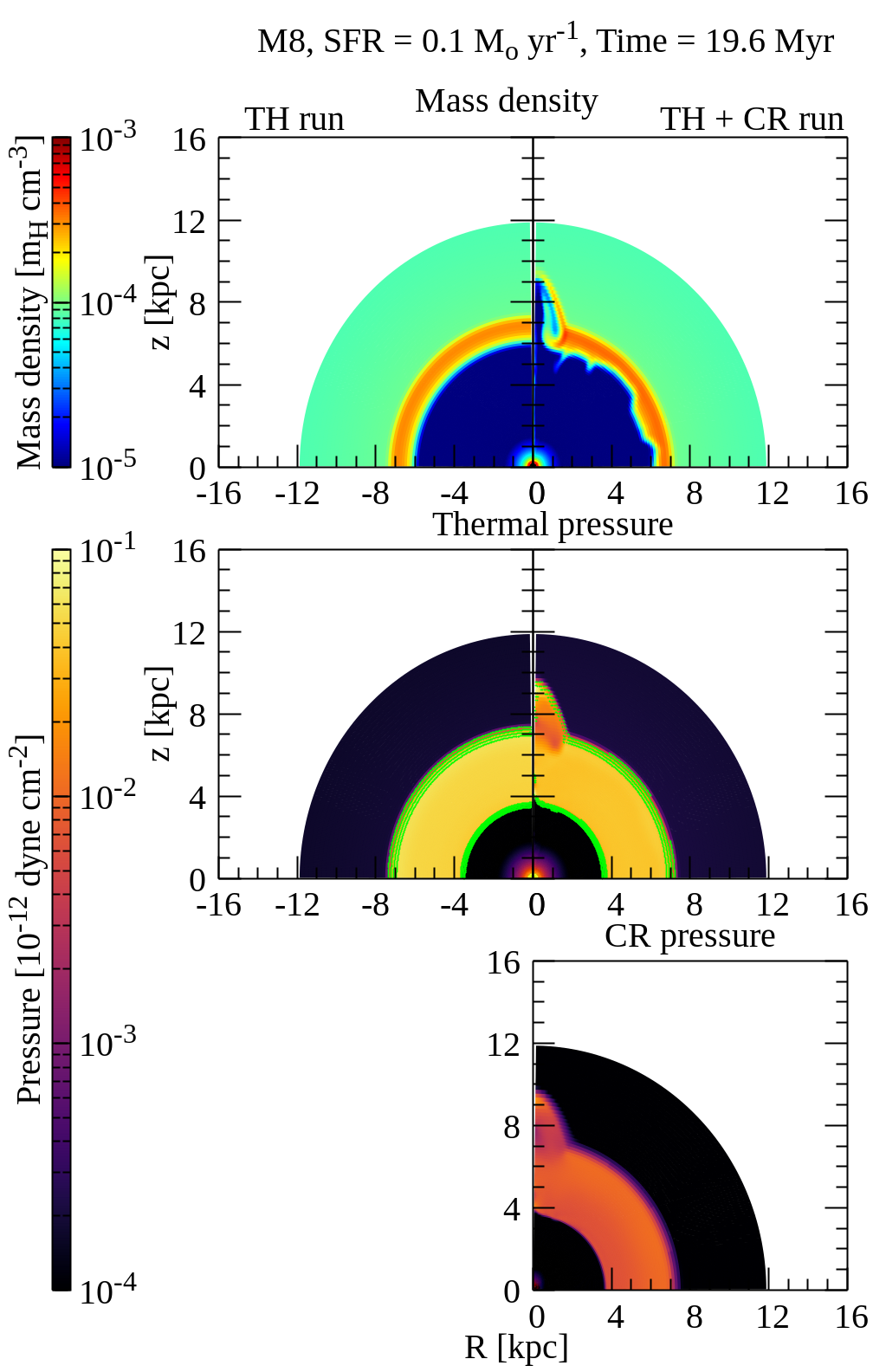}%
\includegraphics[width=2.3in,height= 4.in]{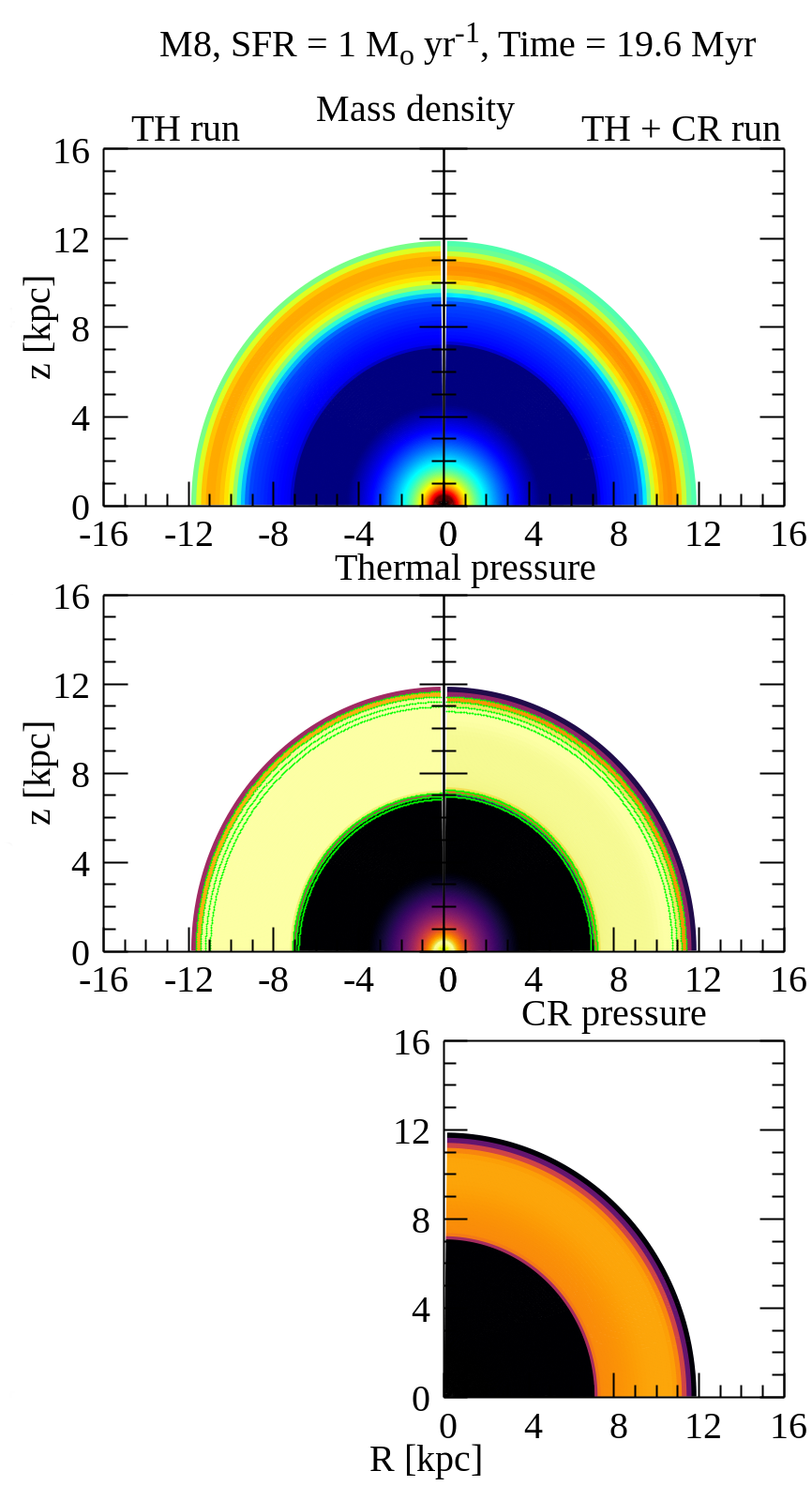}%
\caption{Same as Fig. \ref{fig:M12_compare}, for $M_{\rm vir} = 10^8 M_\odot$} 

\label{fig:M8_compare}
\end{figure*}

In Fig. \ref{fig:M12_compare}, \ref{fig:M11_compare} and \ref{fig:M8_compare}, we show the morphology of outflowing gas for three different halo masses for TH and TH+CR runs. For all runs, we find that the shape of the outflow is nearly spherical in nature. For M8, the shape is expected to be more spherical because, in this case, the galactic disk is much smaller than the outflow length scale and the disk is not included in our simulation. For all cases, the forward and reverse shocks are prominently distinguishable, except for SFR =0.1 $\sfr$ in M12 galaxy. In this case the outflow has an irregular shape and does not show a prominent reverse shock; the locations of shocks are rather distributed inside the bubble. This can be seen from the shock locations, as displayed by green dots in the second rows of Figs. \ref{fig:M12_compare}, \ref{fig:M11_compare} and \ref{fig:M8_compare}. 

For SFR = 0.1 $\sfr$ (first column of Figs. \ref{fig:M12_compare}, \ref{fig:M11_compare}, and \ref{fig:M8_compare}), a comparison of different snapshots among three halos shows that for M12 galaxy, before the outflow breaks out of the central region, an accoustic wave propagates from the central region. This can be understood as follows. 

In our set-up, CRs are initially confined in the galactic disk and the halo has very low CR pressure. Due to the pressure gradient, the CRs start to diffuse out of the disk and give rise to the acoustic wave. This feature is not prominent for SFR = 1 and 10 $\sfr$ because the velocity of the outflow is greater than this acoustic wave generated due to diffusion of CRs, hence the outflow overtakes it. In other words, such waves can be noticed if the diffusion timescale is smaller than the dynamical timescale of the outflowing gas. The diffusion time scale to reach the disk scale height $b$ is given by
\begin{equation}\label{eq:tdiff}
\tau_{\rm diff, CR} \approx \frac{b^2}{6\,D_{\rm cr}} \sim 1\,{\rm Myr} \Bigl ( {b \over 0.4 \, {\rm kpc}} \Bigr )^2 \, D^{-1}_{\rm cr, 28}\,
\end{equation}
where $D_{\rm cr} = 10^{28} D_{\rm cr, 28}$ cm$^2$ s$^{-1}$ is the diffusion coefficient. In case of M11 this acoustic wave initiated by diffusion is not visible because M11 has a puffed up disk compared to M12 (due to the weaker gravity and less steep temperature gradient from the disk to the halo, see Fig. \ref{fig:cooltime} in Appendix A). Therefore, in this case the acoustic wave is initiated at a later time (due to longer diffusion timescale of CRs) and by that time, it is overtaken by the outflow.

\subsection{Outer shock position}\label{sec:fwd_shock}
\begin{figure*}
\includegraphics[width=5in,height=4in]{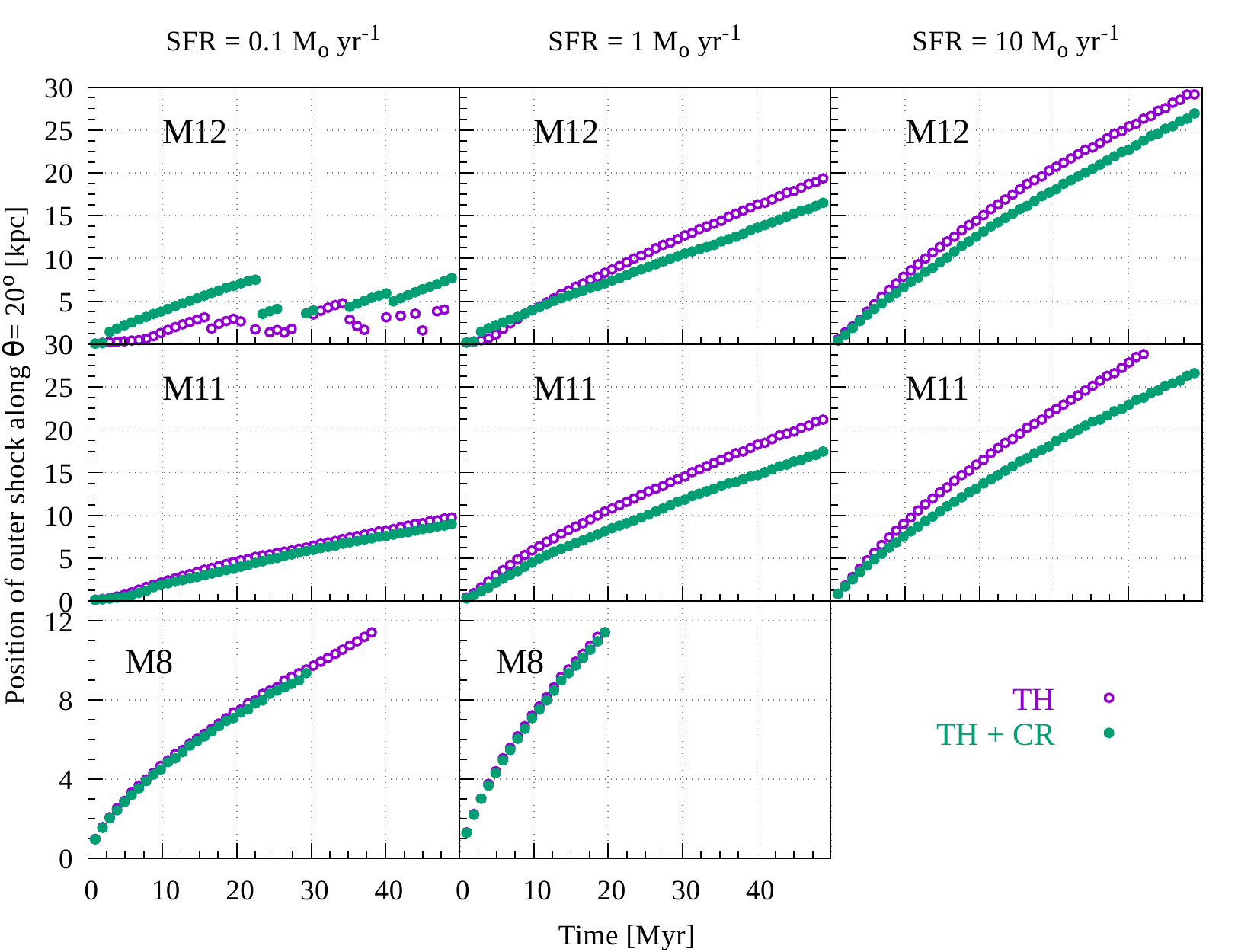}
\caption{Position of forward shock detected at $\theta = 20 \degree$ for M12, M11 and M8 galaxy for star formation rate 0.1, 1 and 10 $\sfr$. The purple dots are for TH run and the green dots are for TH+CR run with diffusion coefficient, $D_{\rm cr}$ = $10^{28} \rm cm^2 s^{-1}$. It is clear that the shock traverses a shorter distance when CR is included since a fraction of the energy within the bubble (region within outer shock) escapes  due to CR diffusion.}
\label{fig:fwd-shock}
\end{figure*}

\begin{figure*}
\includegraphics[width=0.75\textwidth]{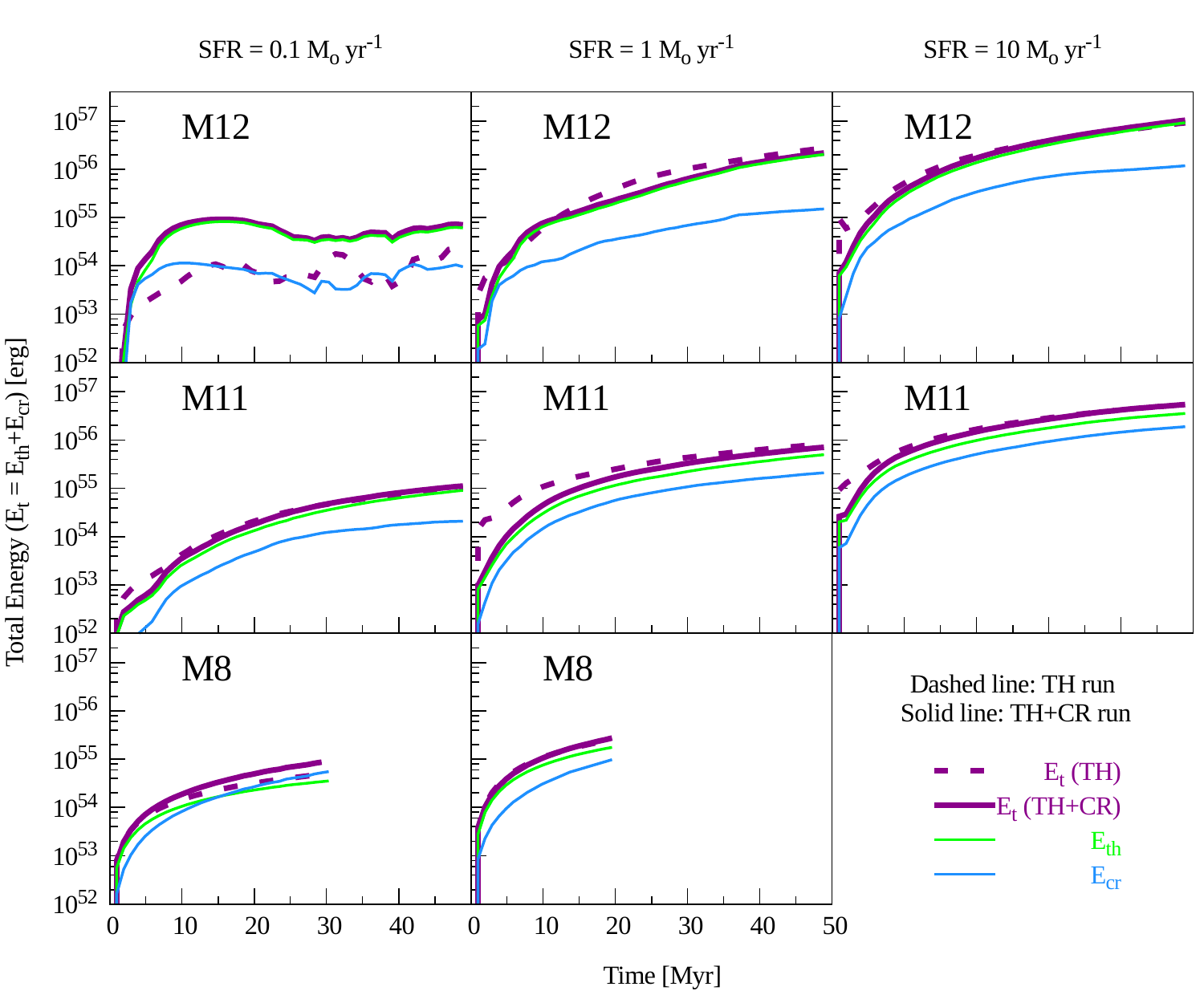}
\caption{Total energy {\it within} the bubble (region within outer shock) is plotted for the TH (dotted purple lines) and TH+CR runs (solid purple lines) for all the cases. The green and blue solid lines show the thermal and relativistic component respectively in each TH+CR run. Comparing the dashed and solid purple lines we notice that, in most of the cases, the total energy within the bubble becomes slightly less when CRs are included, because CR energy leaks out due to diffusion.}
\label{fig:tot_energy}
\end{figure*}

We show the positions of the forward shocks\footnote{One can also choose $\theta = 0 \degree$ to show the positions of the forward shocks, however we notice that it may show some artifacts due to boundary effects.} at $\theta = 20 \degree$ for three different halo masses in Fig. \ref{fig:fwd-shock}. In all panels, purple dots represent TH runs and green dots represent TH+CR runs. The positions of the forward shocks for M12 galaxy with low SFRs are difficult to determine, since they do not have a distinct shock structure w.r.t. CGM gas. From the figure, it can be noticed that for higher SFRs, the forward shock distance is \textit{decreased} by the inclusion of CRs (green dots are located below purple dots). The reasons are illustrated as follows:

Injections of CRs at the shocks develop non-thermal pressure in the downstream region, namely, the shocked-wind  and the shocked-CGM gas. The effective adiabatic index $\gamma$ of the composite gas in these regions is determined by the CR injection fraction at the shock. For a non-zero $w$, the effective $\gamma$  is reduced from  $5/3$ to $(5+3w)/[3(1+w)]$ \citep{Chevalier1983}. Consider two extreme cases, of same amount of energy being either channeled to thermal gas or to CRs. From energy conservation, we then have $3\,p_{\rm cr} V_{\rm cr}={3 \over 2} p_{\rm th} V_{\rm th}$, where $V$ denotes volume, with subscripts referring to either the case of CR or thermal gas. The pressure in the shocked region is a certain fraction of the ram pressure ($\rho_{\rm ambient} v_{\rm shock}^2$), and at a given time, they are nearly same (\citealt{Weaver1977}) in two cases. Therefore $V_{\rm cr}< V_{\rm th}$. In other words, the volume occupied by shocked gas is expected to be smaller if the gas is mostly dominated by CRs. Previous studies (for eg. \citet{Pfrommer2017}, see Fig. 3, top right corner) that investigated blastwave with CR shock acceleration through 3D simulations also reported similar results.
This effect is  enhanced by CR diffusion. Diffusion causes the leakage of CR energy from the bubble, which further reduces the size of bubble.\footnote{The reduced value of effective $\gamma$, and the loss of energy through CR diffusion (as in the case of radiation loss) also implies a slightly higher density jump behind the shock, which we have confirmed from our simulation (see for e.g., Fig. 2 in \citealt{Gupta2018Jan}).} This is why the outer shock distance is decreased in the case of TH+CR.

We further elaborate upon these issues in Section 4.2. We show that, in the case of with-CRs no-diffusion run, the outer shock position is slightly smaller than without CR runs. When diffusion is turned on, the difference becomes much larger and the outer shock travels to a shorter distance.
In Fig. \ref{fig:tot_energy}, we have shown the total energy ($E_{\rm t}$) within the bubble (i.e., within forward shock) as a function of time. In case of TH+CR runs the total energy includes the thermal as well as CR energy, i.e. $E_{\rm t} = E_{\rm th} + E_{\rm cr}$ and we have shown these two components separately. Fig. \ref{fig:tot_energy} shows that in case of TH+CR run, for most of the cases, CR energy is less than the thermal energy and it is expected since our injection prescription assumes $ w = p_{\rm cr}/(p_{\rm th}+p_{\rm cr}) = 0.2$ (i.e., CR energy density fraction $e_{\rm cr}/(e_{\rm th}+e_{\rm cr}) = 2w/(1+w)=1/3$). Comparison of total energy  (purple curves) between TH and TH+CR runs shows that at early epochs total energy is slightly lower in the case of TH+CR runs, because of CR diffusion. However, this difference decreases as bubbles become larger, because CR diffusion is less effective at larger distances (due to the increase in CR diffusion timescale, which is $\propto (distance)^{2}$, see e.g. Eq.\ref{eq:tdiff}). 

For M12 galaxy, SFR =  0.1 $\sfr$, total energy within the bubble is larger in the TH+CR run compared to the purely thermal run because, here, the forward shock is detected at the position of the acoustic wave. As a result, thermal energy contribution from the CGM is also included in this analysis.

For M8 galaxy and SFR = 0.1 $\sfr$, we notice that the CR energy within the bubble increases with time and after a certain time ($\approx 20$ Myr) it becomes larger than the thermal energy, resulting in an increase of total energy in TH+CR run compared to TH run (purple solid curve above the dashed curve). This is an artifact of boundary effect near the pole for this case, as can be seen in the left column of Fig. \ref{fig:M8_compare}.

\subsection{Mass loading factor}\label{sec:mload}
\begin{figure}
\centering
\includegraphics[width=8.3cm]{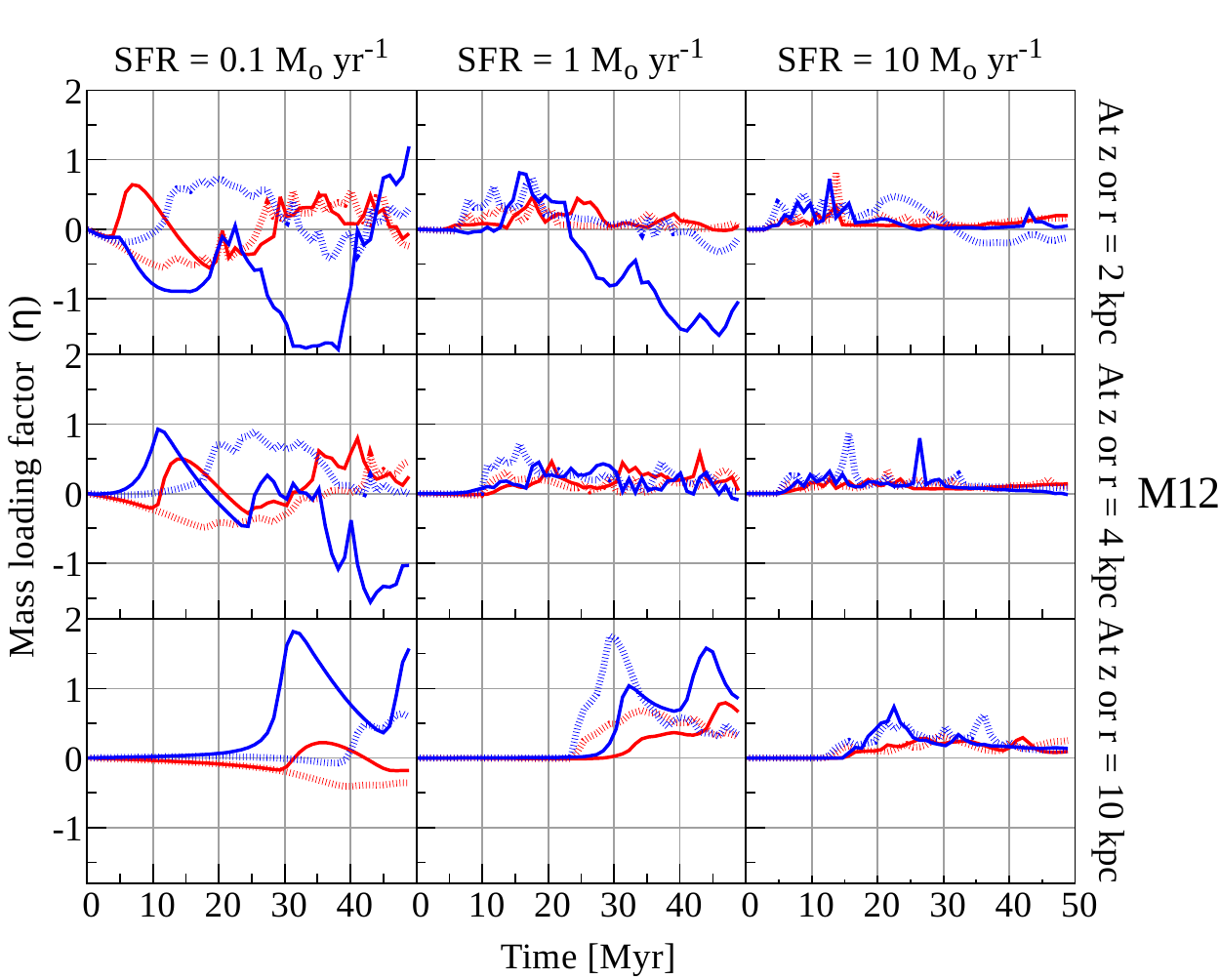}
\includegraphics[width=8.3cm]{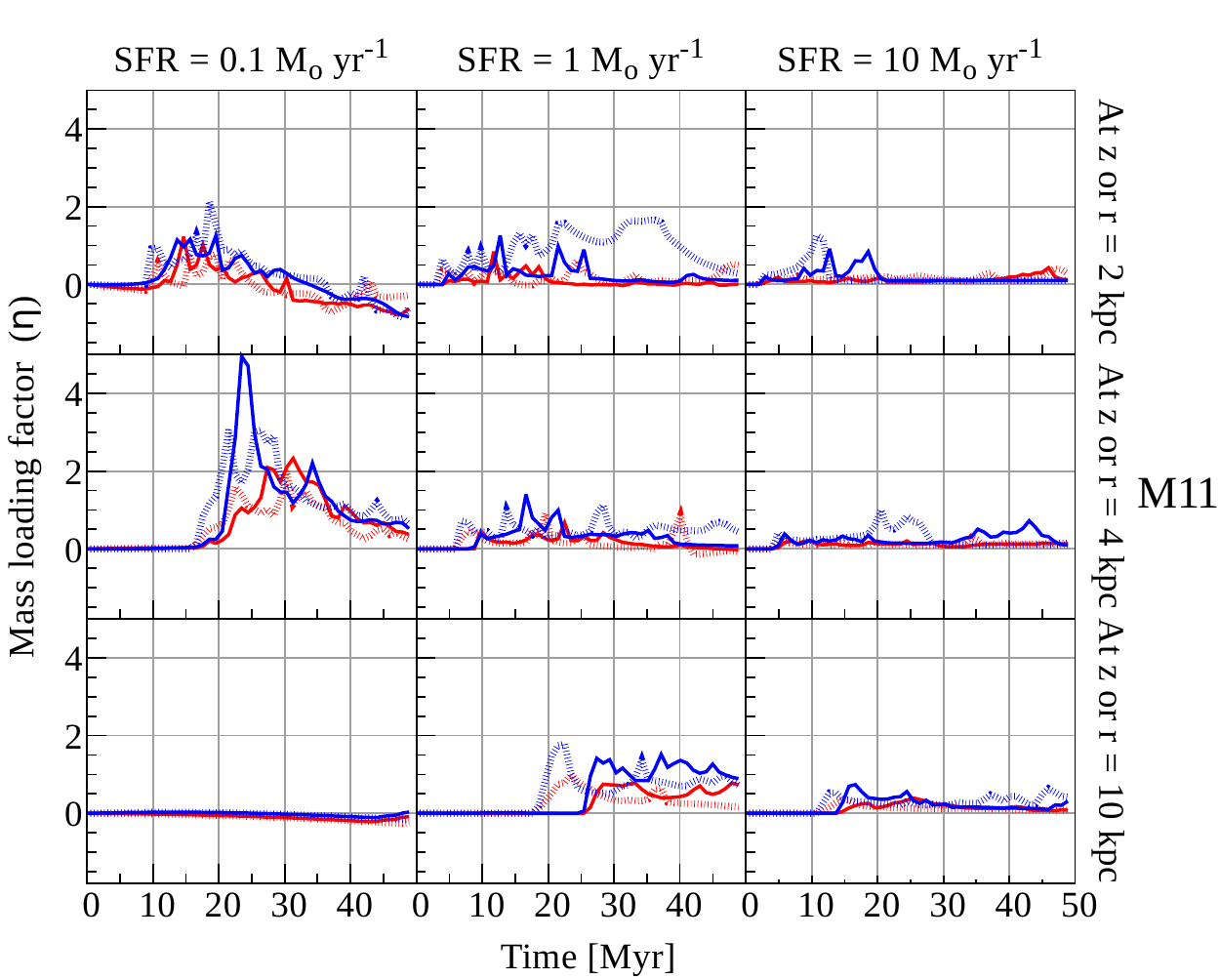}
\includegraphics[width=8.3cm]{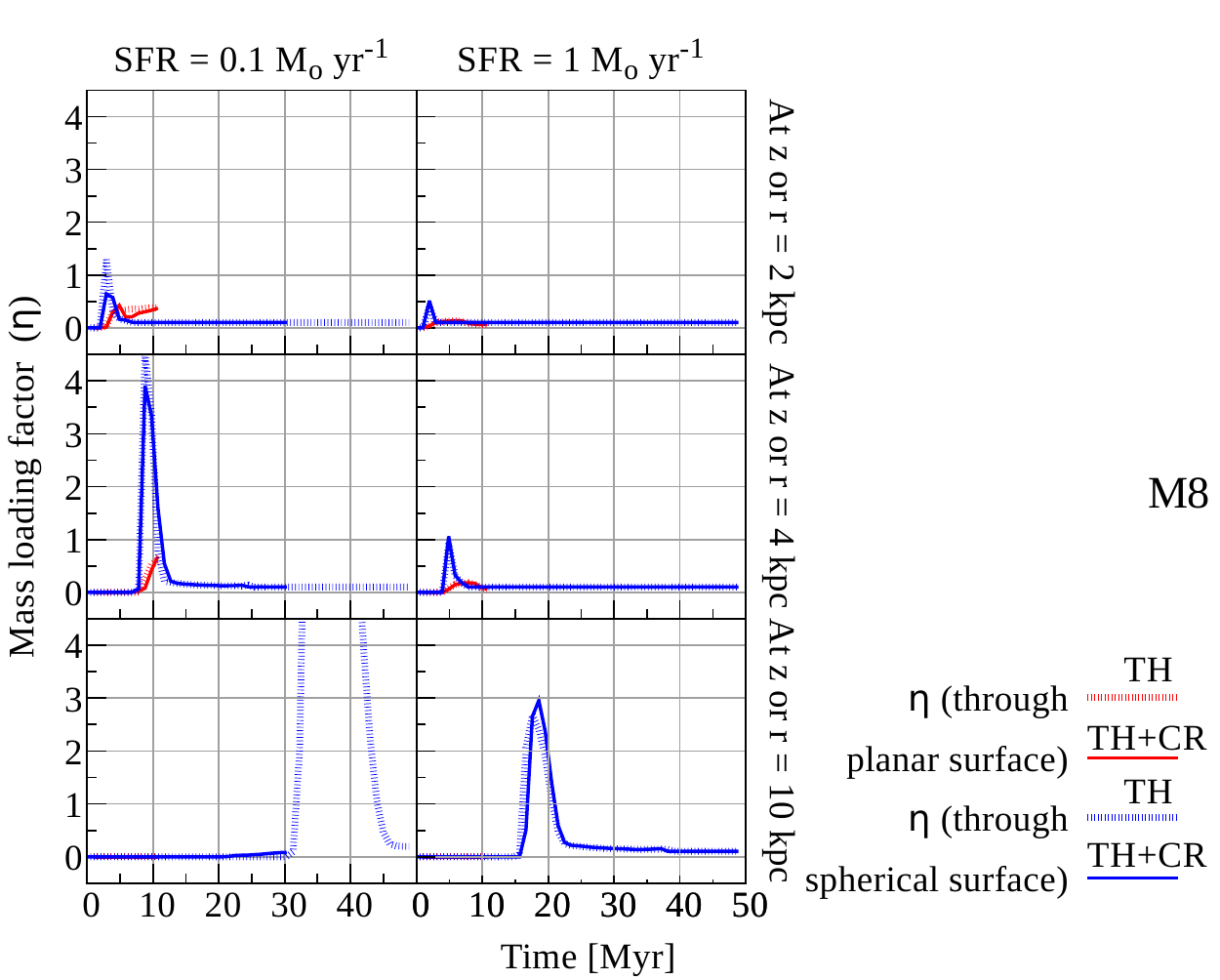} 
\caption{Time evolution of mass loading factor ($\eta$), calculated in two different ways, are shown for all the cases. The first method (red lines) involves the calculation of mass outflow rate through planes at a height of 2, 4 and 10 kpc from the galactic disk and the second method (blue lines) involves the calculation of mass outflow rate through spherical surfaces at radius 2, 4 and 10 kpc from the center. In each case, the dashed line is for TH runs and the solid lines are for TH+CR runs. We see no significant difference between the dashed and solid lines. Galaxies with higher mass and lower SFR has more negative values of $\eta$ owing to the infall of gas due to gravity.}
\label{fig:mload}
\end{figure}

Mass loading factor ($\eta$) is defined as the ratio of the mass outflow rate through a given surface to the SFR at a given epoch. Despite a simple definition, there is disparity between the way  mass loading factor is determined in simulations and in observations. One main difference is in the definition of the surface area through which mass outflow rate is calculated. For example, \citet{Booth2013} defined the surface as a plane at $20$ kpc above the disk and \citet{Jacob2018} used a cylinder of different radii and of height reaching up to the virial radius.
However, observers define the mass outflow rate as $\approx \Omega N \, r \, v \, m_p$, where $\Omega$ signifies the total solid angle considered, $N$ is the column density of material along the line of sight,  $r$  is a characteristic distance of the location of most of the outflowing material and $v$ is the outflow speed \citep{Heckman2015}. Such differences make it difficult to interpret the results.

In order to compare our results with previous works, we use two different definitions of mass loading factor, as described below.

\subsubsection{Outflow rate across a plane}\label{sec:mload_plane}
In this case, we estimate the mass outflow rate through a plane parallel to the galactic disk. To calculate the mass loading factor from our simulation, we define three planes at distances $z=2\,,4$ and $10$ kpc above and below the galactic disk of radius\footnote{We have checked that the choice of radius ($R_{p}$) does not affect the calculations as long as the extent of the outflowging gas does not exceed this radius. For M8, this constraint restricts our analysis to an epoch beyond which the outflow becomes larger than $5$ kpc.} $R_{p}=15$ kpc for M11 and M12, and $R_{p}=5$ kpc for M8. The mass loading factor $\eta$ has been calculated using,
\begin{equation}
\eta\equiv \frac{\dot{M}_{\rm load}}{{\rm SFR}}= 2\times \frac{\Bigl [\int_{R=0}^{R_{p}} \hat{z}\cdot d{\bf A}\, \rho\,  v_{\rm z}\Bigr ]}{\rm SFR}\, ,
\end{equation}
where $d\mathbf{A} =  \hat{z}\, 2\pi R\,dR$ is a differential area element in the plane and $ v_{\rm z}$ is the vertical component of gas velocity. The factor $2$ takes into account the contributions in mass flux both above and below the galactic disk. 

We show the mass loading factor ($\eta$) as a function of time for three different galaxies in Fig. \ref{fig:mload} by red curves (solid: TH+CR run, dashed: TH run). The curves show that, in the early epochs, $\eta$ rises to a peak value and thereafter it becomes small. In some cases, it also shows negative values because of infalling gas. The peak(s) in the evolution of $\eta$ is due to the crossing of the swept-up dense shocked gas near the contact discontinuity. After this epoch, the value of $\eta\sim 0.1$, which comes from the gas in the free wind region. In this region, the mass loading factor is  $\approx \dot{M}_{\rm inj}/{\rm SFR} \simeq 0.1$ because we have taken $\dot{M}_{\rm inj}$ as $10\%$ of the SFR\footnote{We can also analytically estimate the mass loading factor from the self-similar solution, calculating the contribution of the shocked gas (shell) and free wind separately (see Appendix \ref{app:analytical_mload}).}. 

Comparing the solid (TH+CR) and dashed (TH) lines, we find no significant difference in the magnitude of $\eta$ in case of SFR $= 1$ and $10$ $\sfr$. For M12 galaxy, SFR $= 0.1$ $\sfr$, the time variation of $\eta$ has a distinct feature  for the run with CR, displaying a peak that  propagates with time to higher $z$. This is the signature of the acoustic disturbance already described in the Section \ref{sec:morphology}.

\subsubsection{Outflow across a spherical surface}\label{sec:mload_shell}
Next, we estimate the mass outflow rate through a spherical surface of radius $r$ centered at $r=0$ as,
 \begin{equation}
 \eta\equiv {\dot{M}_{\rm load} \over{\rm SFR}}= \frac{4 \pi r^2}{\rm SFR}  \int_0 ^{\pi/2}  \rho v_r \sin \theta \, d\theta \, 
  \end{equation}
  where $v_r$ is the radial component of the gas velocity. We show the mass loading factor as a function of time for different galaxies with blue lines (solid: TH+CR run, dashed: TH run, as used earlier) in Fig. \ref{fig:mload}. 
  
The curves show that the evolution of mass loading factors qualitatively remains similar to the planar case (Section \ref{sec:mload_plane}). However the peak value in this case is larger than that in the previous case. Since the morphology of the outflowing gas is spherical, considering a spherical surface results in a larger mass outflow rate.

We note that mass loading factor is prominently negative for M12 and SFR =0.1 $\sfr$, for z =$2$ kpc and $r = 2$ kpc. One can understand this by considering two relevant speeds, namely, the outflow speed and the escape speed. The outflow speed is roughly $100$ km s$^{-1}$, compared to the circular speed of $150$ km s$^{-1}$ of such a galaxy and also the escape speed of the dark matter halo (Fig. 2 of \citealt{Sharma2012}). This results in the infall of cold gas clumps produced in the shocked gas layer, producing negative values of $\eta$. In other words, galaxies with higher mass and lower SFR are likely to exhibit negative values of $\eta$. Fig. \ref{fig:mload} also shows negative values in some other cases at later times, which essentially comes from the infall of cold clumps produced near the base of the outflow  for the Kelvin-Helmholtz instability due to shear.

We notice in Fig. \ref{fig:mload} that $\eta$ tends to decrease with increasing SFR, except in the high-mass galaxy low-SFR cases, where $\eta$ often becomes negative due to the effect of stronger gravity. This effect is most prominent for M8 galaxy where gravity effects are the least. From the analytical solution by \citet{Weaver1977}, the position ($r$) of the outflow at time ($t$) is estimated as $r \propto (L t^3/ \rho)^{1/5}$ where $L$ is the luminosity of the source and $\rho$ is the density of ambient gas. Hence the velocity ($v$) of the outflow when it is at distance $r$ is $v \propto (L/\rho)^{1/3} r^{-2/3}$. The mass outflow rate ($\dot{M}_{\rm out}$) at $r$ would be proportional to the mass swept up by the outflow ($ \propto r^3 \rho$, does not change with $L$ ) and the outflow velocity $v$. Hence $\eta = (\dot{M}_{\rm out}/{\rm SFR}) \propto L^{-2/3}$ or $\eta \propto {\rm SFR}^{-2/3}$ since SFR is proportional to the source luminosity. The peaks in Fig. \ref{fig:mload} for M8 galaxy is consistent with this expected scaling.
\begin{figure*}
\centering
\includegraphics[width=0.5\textwidth,height= 5.1cm]{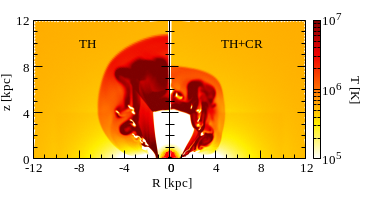}%
\includegraphics[width=0.5\textwidth]{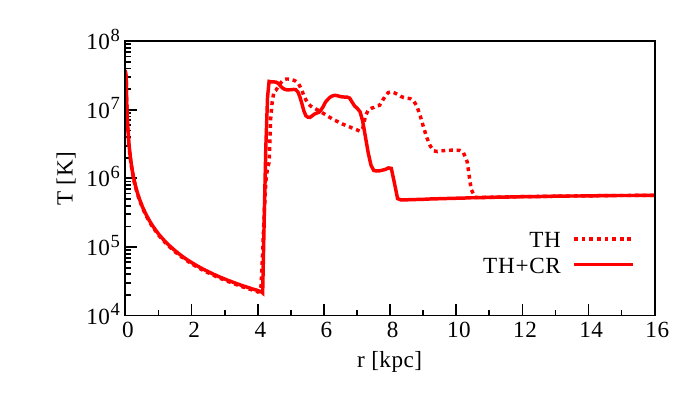}
\caption{For the fiducial case (M11 galaxy, SFR = 1 $\sfr$, Time $\approx$ 20 Myr) both the 2D temperature plot and the 1D temperature profile at $\theta = 20\degree$ show that shocked CGM in the outflow has a lower temperature when CRs are included compared to purely thermally driven outflow.}%
\label{fig:temp}
\end{figure*}

\begin{figure*}
\centering
\includegraphics[width=5.5in,height=4.5in]{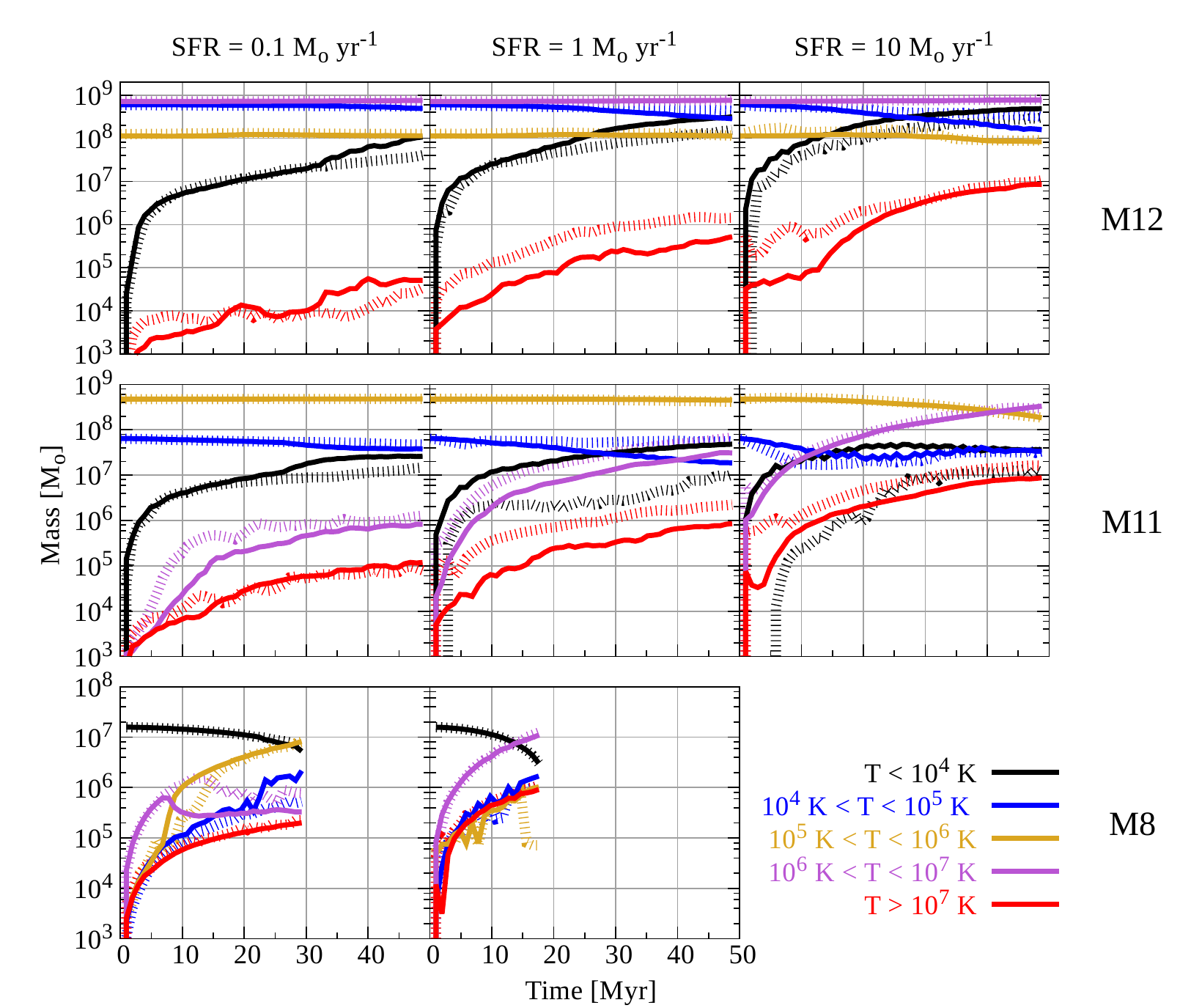}
\caption{Gas mass contained in the simulation box are shown with dashed (TH) and solid (TH + CR) lines of different colours for different temperature ranges. It shows that gradually amount of very cold gas (T $< 10^{4}$ K) becomes more in the cases where CRs are included, compared to purely thermal models.}%
\label{fig:coolgas-dist}
\end{figure*}

\begin{figure}
\centering
\includegraphics[width=0.5\textwidth, left]{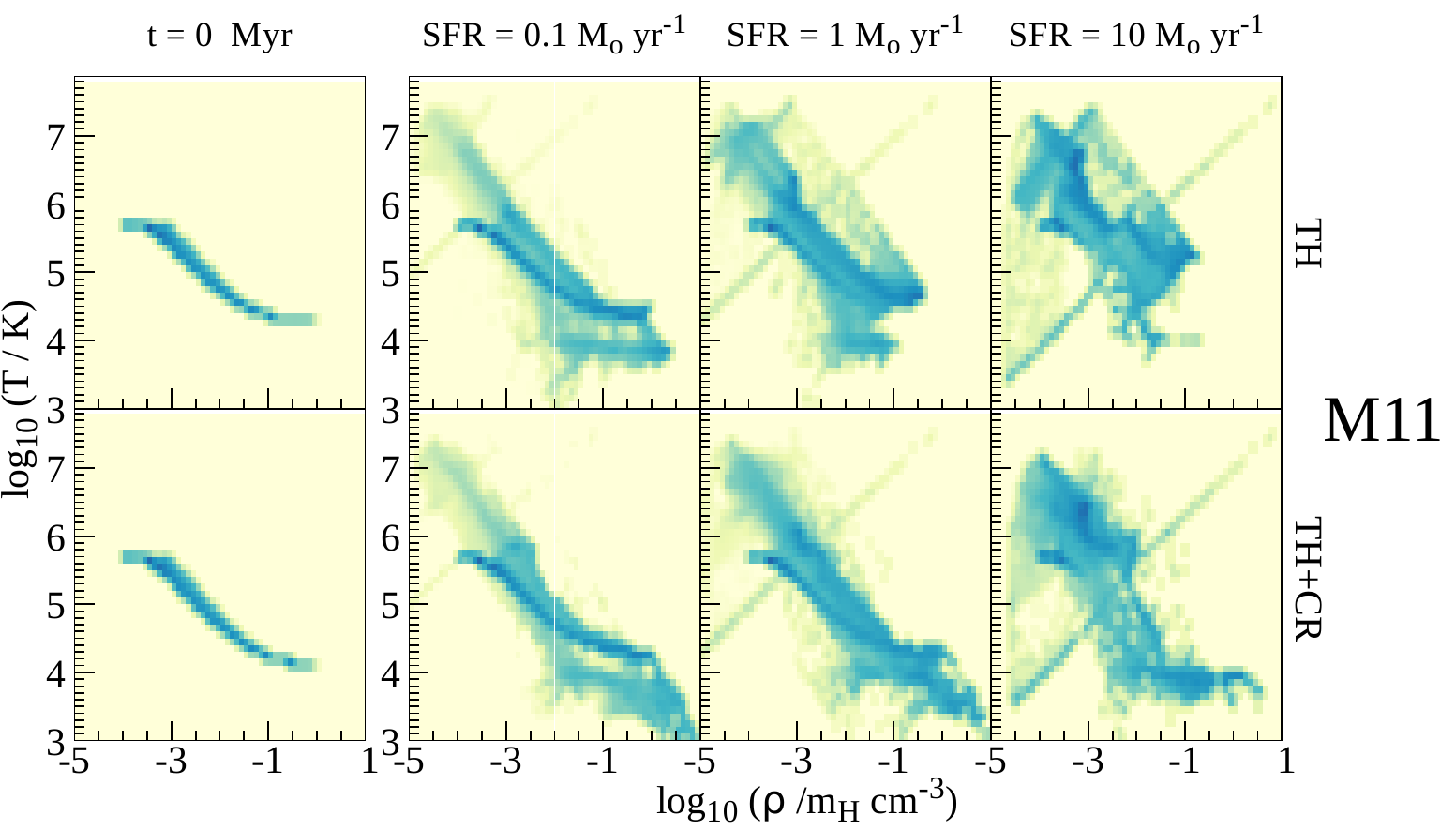}
\includegraphics[width=0.5\textwidth ,left]{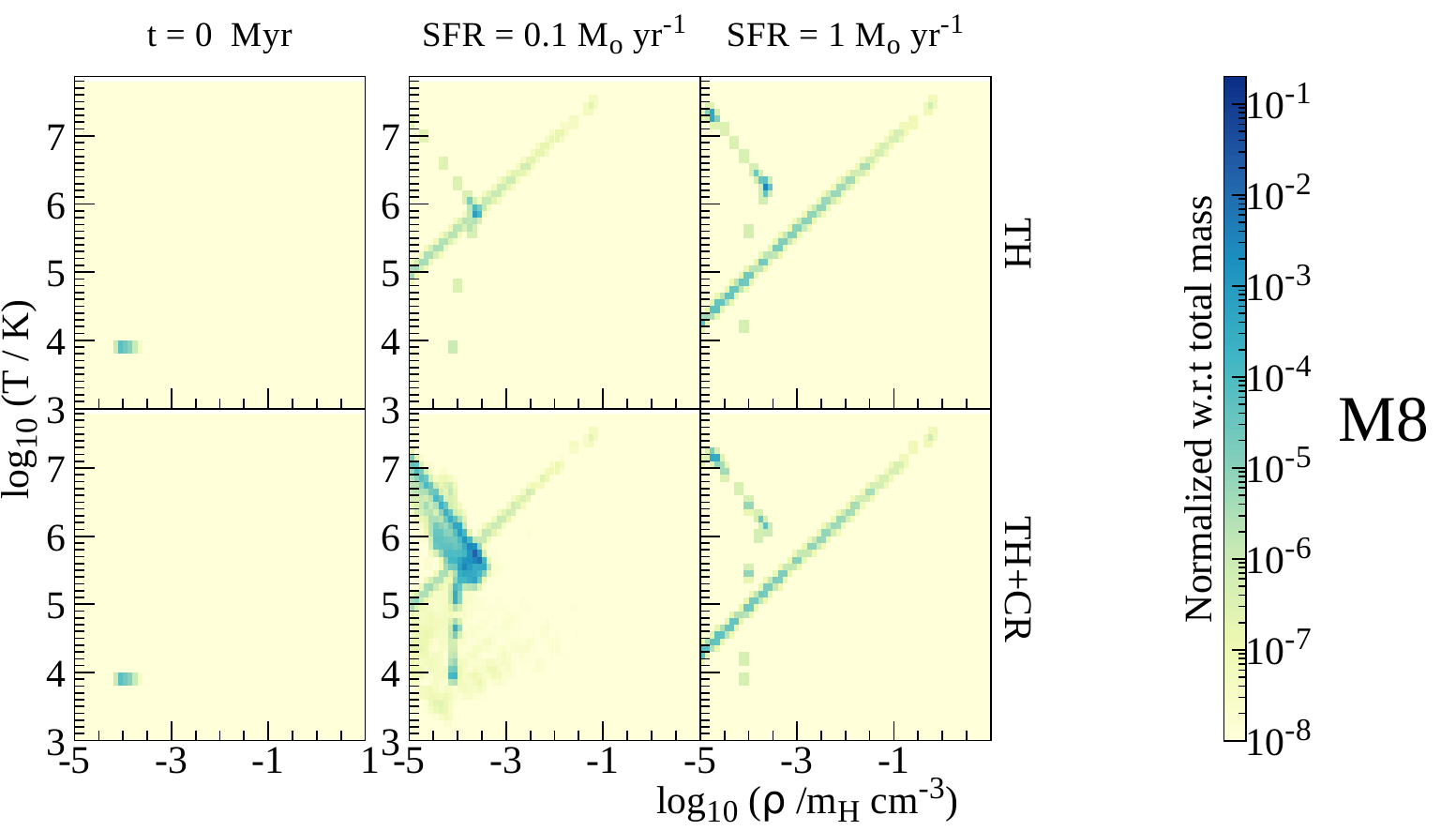}
\caption{The 2-D probability distribution of gas mass fraction (normalized to total mass in the simulation box) within specified density and temperature bins at $\approx 20$ Myr for M11 and M8 galaxy. The left most columns for each panel show the initial mass probability distribution. This figure suggests enhanced cooling of shocked gas when CRs are included, as indicated by the portions with $T \le 10^4$ K in TH+CR runs compared to TH runs.}
\label{fig:den-temp-dist}
\end{figure} 
 
\subsection{Temperature distributions}\label{sec:temp-dist}

It has been claimed in previous studies that the shocked gas in a CR driven outflow is at a lower temperature compared to a purely thermally driven outflow. Fig. \ref{fig:temp} shows that for the fiducial case, our simulation results are in accordance with this findings. The temperature of the shocked gas at the outer shock is $\sim 1.7$ times lower in TH+CR run.

To undestand the role of CRs in multiphase structure of the outflow the temperature distribution of gas has been used in different studies. We study the time evolution of the total mass content in five temperature bins. The temperature bins represent different phases of the gas:
\begin{itemize}
\item very cold ($T<10^4$ K) 
\item cold ($10^4$ K $ <T<10^5$ K)
\item warm ($10^5$ K$ <T<10^6$ K)
\item hot ($10^6$ K $<T<10^7$ K)
\item very hot ($T >10^7$ K)
\end{itemize} 

The time evolution of the total mass in each temperature bin is shown in Fig. \ref{fig:coolgas-dist}. The curves show that the simulated galaxies M12 and M11 initially have almost negligible gas content in very cold ($T<10^4$ K) and very hot ($T>10^7$ K) range. However, the gas mass in these two phases gradually grow with time. The increase in very cold gas content suggests cooling due to adiabatic expansion of gas, whereas the increase in very hot gas content suggests the increase of mass content in the shocked wind region with time. For M12, the halo gas temperature is $3 \times 10^6$ K and for M11 it is $6\times10^5$ K, and so the gas mass in these bins for the respective galaxies remains almost constant since the maximum contribution in these ranges come from the halo gas. In both M12 and M11, the cold ($10^4$K$<T<10^5$K) gas content decreases gradually with time which suggests the mass loss from the disk due to outflow. For M8, since all the gas is initially at $10^4$ K, the gas mass in all phases increases with time as expected, except the  very cold ($T<10^4$ K) gas.

Comparing the results for runs with and without CRs, we find that the amount of very hot ($>10^7$ K) gas in TH+CR runs is $2\hbox{--}4$ times lower compared to TH runs in both M12 and M11. This is expected from the consideration that a fraction of the total energy in shocked gas is imparted to CRs, thereby reducing the thermal energy density there, consequently reducing the temperature. This has been known since the calculations by \citet{Chevalier1983} for blast and driven waves with cosmic rays, where it was shown that the temperature of the hot gas behind the shock is reduced in the presence of CRs (see also \citealt{Gupta2018Jan}). Since a reduction of temperature increases the radiative cooling rate in this temperature range, the presence of CRs also increases cooling and the amount of cold gas. This is borne out by the fact that in our simulation, the amount of very cold ($T<10^4$ K) gas in TH+CR runs is $2\hbox{--}5$ times higher than runs without CRs. This finding of increase of cold gas in the presence of CRs is consistent with previous studies.


To get a better idea of the multiphase structure of the outflow, we show a 2-D probability distribution of gas mass in the simulation box within specified density and temperature bins [$\rm \Delta log_{10}(T/{\rm K}) =0.1$ and $\rm \Delta log_{10}(\rho/{\rm m_{\rm H}\,cm^{-3}}) =0.1$] at $\approx 20$ Myr in Fig. \ref{fig:den-temp-dist}. Gas mass in each density and temperature bin has been normalized by the total gas mass within the simulation box. In the initial distribution of M11 galaxy, the bright spot in the low density and high temperature ($\sim 10^6$ K) region represents the hot and tenuous halo gas. The high density and low temperature ($\sim 10^4$  K) region is the signature of the dense and cold disk gas. The gas that falls between these two regions follow a straight line with negative slope ($\approx -0.8$) in the $\rho - \rm T$ diagram. This negative slope which approximately follows the power law, $T \propto \rho^{-1}$ indicates that the initial density distribution is in pressure equilibrium\footnote{The slope is not exactly -1 since gravitational force is also included in setting up the initial hydrostatic equilibrium.}, as expected for a stable galaxy set-up. For M8 galaxy, such a region is absent in the initial distribution since there does not exist a wide distribution of density as well as temperature. Most of the halo gas has density $\sim 10^{-4}  m_{\rm H} \rm {cm}^{-3}$ and temperature $\sim 10^4$ K.

At a later time, at $\approx 20$ Myr, a fraction of the CGM (halo) gas reaches higher temperature and lower density compared to their initial states, arising from the shocked wind region of the outflow. Another feature to be noticed is that an amount of gas is accumulated towards the bottom right corner of the diagram, i.e., at a high density and low temperature region. This is the shocked gas near the weak outer shock in the disk.

For SFR = $1$ and $10$ $\sfr$, another interesting aspect is that, for M11 galaxy, there is an accumulation of gas that shows up along a straight line with positive slope in the $\rho - \rm T$ diagram. This line is a characteristic of adiabatic expansion in the free wind region of the outflow. We do not see this feature for SFR = 0.1 $\sfr$  because the wind is not powerful enough and as a result, the outflow may not have a distinct free wind region, as previously shown in Fig.  \ref{fig:M11_compare}. However, for M8 galaxy this feature is noticeable even for low SFR because in this case, outflow is strong enough due to weaker gravity of the halo and the absence of a thick disk compared to M11 .

Comparing the results for TH and TH+CR runs for M11, we find that the temperature of the shocked disk gas (right bottom corner) is lower by a factor $\sim 2$. These results are also valid for M12, albeit these effects are more prominent for M11 galaxy.

\section{Effect of long duration outflow evolution}\label{sec:long_dur}
In previous sections, we have presented results by focusing on first $\sim 50$ Myr of galactic outflows. We would like to emphasize that our idealized simulation in which the star formation is not coupled to any feedback processes is a controlled experiment, and the aim is to determine the role of CRs by comparing our findings with previously published results. In order to understand the long term evolution, we extend the simulation of our fiducial galaxy (M11) until 210 Myr. Since, SFR cannot be kept constant for such a long duration, we consider a period of star formation events with the period being 30 Myr. In other words, star formation is kept uniform between $0\hbox{--}30, \, 60\hbox{--}90, \,120\hbox{--}150, \,180\hbox{--}210$ Myr, and it is switched off in the intermediate periods. As an alternative to using two different free parameters, we assume the period of star formation and the quiescent period both to be 30 Myr, being motivated by \citet{Booth2013}. This allows us to study the structure and evolution of sequential outflows triggered by periods of star formation, which produces numerous shocks.

We find that each star formation event gives rise to a shock structure. The subsequent shocks produced after the first event of star formation, travel faster than the previous shocks since they travel in a comparatively rarefied medium. It can be observed that after sufficient amount of time, the subsequent shocks catch up with the first shock produced. This phenomenon has been previously noticed by \citet{Tang2005}. However, as found in the short duration runs, here also we find that the shocks travel to a shorter distance in the presence of CRs.

\subsection{Mass loading in the long term evolution}\label{mload_longterm}
\begin{figure}
\centering
\includegraphics[width=0.48\textwidth]{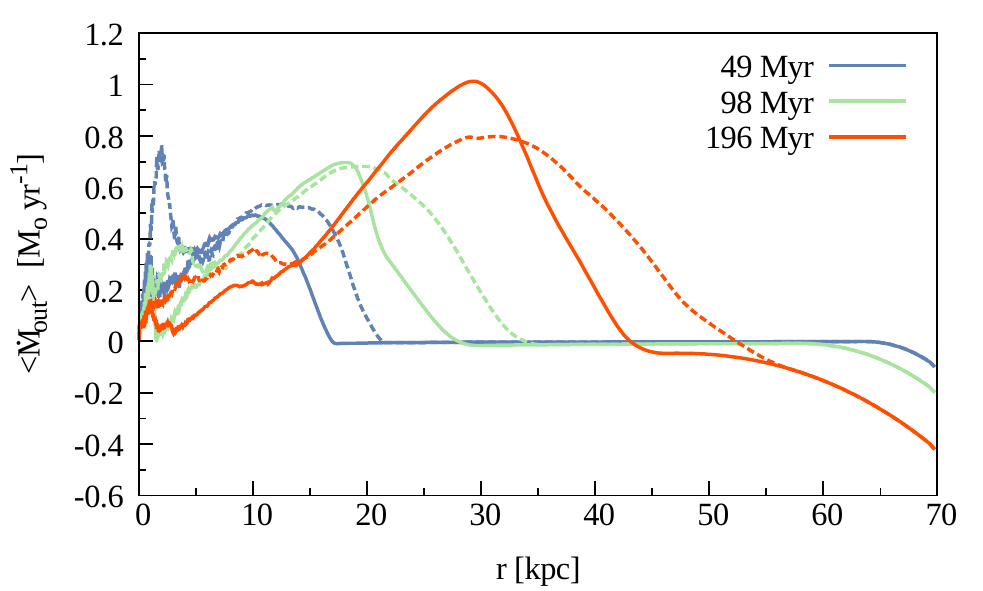}
\caption{Time averaged mass outflow rate, $\langle \dot{M}_{out}\rangle$, as a function of distance for with (solid curves) and without (dashed curves) CR  runs. Three colours denote the epoch up to which the time average is done (see Eqn. \ref{eqn:mload_long} ). Figure shows that with and without CR runs give a similar evolution of $\langle \dot{M}_{out}\rangle$.}
\label{fig:mload_time}
\end{figure}

To study the mass loading in the long-duration simulation with periodic star formation, we calculate the time averaged mass outflow rate instead of the previously used mass loading factor. At a particular time (say, $T_{\rm dy}$) we show the time averaged mass outflow rate through each spherical surface of radius $r$,
\begin{eqnarray}\label{eqn:mload_long}
\langle \dot{M}_{out}(r)\rangle &=& \frac{1}{T_{\rm dy}}\int_{0}^{T_{\rm dy}}  \dot{M}_{out}(r,t) \: dt \nonumber \\
&= &\frac{4 \pi r^2}{T_{\rm dy}}\int_{0}^{T_{\rm dy}} \int_{0}^{\pi/2} \rho v_r \: sin\theta \: d\theta  \: dt
\end{eqnarray}
in Fig.\ref{fig:mload_time}. The figure shows that the mass loading gradually increases with time as the outflow progresses outwards and accumulates more mass. After the outflow crosses the region where most of the mass resides within that timecale, the mass outflow rate starts to decline and gradually approaches zero. At large radii, the negative values of mass loading is a numerical artifact, which arises because of the logarithmic grid that makes the radial width of the computational cells large near the outer boundary. We have confirmed that this artifact can be reduced by extending box size, and it does not affect our results.

Comparing the dashed (TH) and solid (TH+CR) curves of Fig.\ref{fig:mload_time} we note that mass loading is almost independent of whether the outflow is driven by purely thermal gas or a composition of thermal and CR gas. However, we notice that $\langle \dot{M}_{out}(r)\rangle$ has a sharper peak when CRs are considered and at later time, 100 Myr onwards, the peak is slightly higher ($\sim$ 1.3 times) compared to TH case. This analysis justifies our previous results on mass loading and it verifies that the findings are robust as they are also valid in case of multiple star formation episodes in a longer time-scale.

\begin{figure*}
\centering
\vspace{-1. cm}
\includegraphics[width=\textwidth]{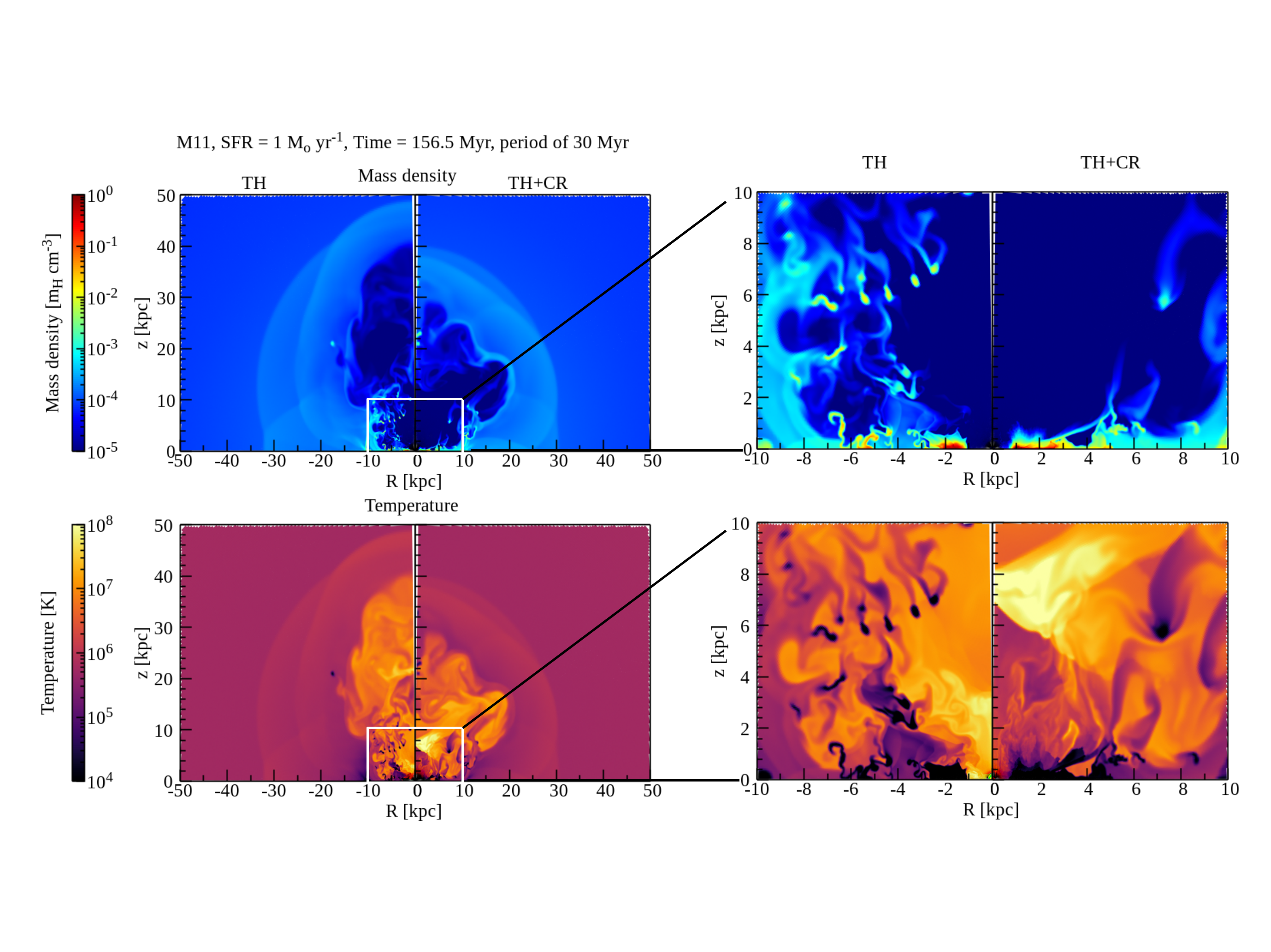}
\vspace{-2. cm}
\caption{Density and temperature distribution of M11 galaxy at 156.5 Myr in the long duration simulation with periodic star formation. The outer shock produced in the first star formation event evolve independently, however the shocks produced in the subsequent star formation events travel in a comparatively rarefied medium and can catch up with the first shock. In the zoomed in version (shown on the right) of the density and temperature distribution one can see \textit{more} high density and low temperature gas clumps in TH case compared to the TH+CR run. (for details, see Section \ref{sec:cold_clumps})}
\label{fig:cold-clump}
\end{figure*}

\subsection{Nature of cold clumps and the role of CR feedback}\label{sec:cold_clumps}
The long duration simulation with periods of star formation allows us to study the formation of cold clumps due to thermal instability in the presence or absence of CRs. This investigation leads to an important insight to the basic role of CRs in galactic outflows. It has been reported (e.g, \citealt{Booth2013, Dashyan2020}) that the presence of CRs can produce a greater feedback effect, which has been shown through an increase in the mass loading factor and a suppression of SFR. As we have found that for time-independent SFR, the mass loading does not show a significant difference between TH and TH+CR runs, we need to study the formation of cold clumps in these two cases. The formation of high density and low temperature gas clumps and their infall on the galactic disk can increase the SFR of the galaxy and thereby can decrease the efficiency of feedback. 

To illustrate this, we show the snapshot of density and temperature distribution at 156.5 Myr in Figure \ref{fig:cold-clump}, which refers to an epoch of 6.5 Myr after the most recent burst of star formation is over (at 150 Myr). It is seen that numerous cold clumps (with temperature $\sim 10^4$ K) are formed in the TH run, while they are absent in TH+CR run. We can understand the inhibition of clump formation by CRs in the following way. The minimum size of cold clumps is given by $c_{\rm s} t_{\rm cool}$ \citep{McCourt2018}, where $c_{\rm s}$ is the sound speed and $t_{\rm cool}$ is the cooling time of the cold gas. If we consider the cooling time of the ambient gas, this will lead to a larger size, but here we are more interested in the {\it lower limit}. On the other hand, while the gas cools, the diffusion of CRs tries to wash out perturbation unless it is larger than  $\ge \sqrt{6D_{\rm cr}t}$, where $t$ corresponds to $t_{\rm cool}$. This is similar to the damping of small perturbation by the streaming of hot dark matter in cosmological context. Therefore the formation and growth of cold clumps requires $c_{\rm s} t_{\rm cool} \ge \sqrt{6D_{\rm cr}t_{\rm cool}}$. This leads to  a lower limit on the cooling time scale
\be
t_{\rm cool} \ge {6D_{\rm cr} \over c_{\rm s}^2} \,.
\ee
Using $D_{\rm cr}\sim 10^{28}$ cm$^2$ s$^{-1}$ and the sound speed of a $10^6$ K gas ($c_s \sim 10^7$ cm s$^{-1}$), this gives a lower limit of $\sim 20$ Myr on the cooling time scale. 
 For a $10^6$ K gas with solar metallicity,
the cooling rate is $\Lambda_{\rm N}\approx 10^{-22}$ erg cm$^3$ s$^{-1}$, and a cooling time scale of $20$ Myr implies an upper limit on the clump density $\le 1.5k_{\rm B}T/\Lambda_{\rm N} t_{\rm cool} = 3 \times 10^{-3}$ cm$^{-3}$. The basic point is that in the presence of CR diffusion, clumps can only form with very low density, and they may not have high density contrast with surroundings.
Since the period between two bursts of star formation here is 30 Myr, if clouds take a long ($\ge 20$ Myr) time to form, then they are likely to be crushed and evaporated by the next outgoing shock, instead of raining down on the galaxy. The inhibition of thermal instability in the presence of CR is therefore a likely explanation of the greater feedback effect of CR found in previous works.

In addition to the above effect, we notice that the shape of the free wind region of outflow changes due to diffusion. Left panel of Fig.\ref{fig:cold-clump} shows that the shape of the free wind is more elongated along $z$ direction when the outflow is driven by purely thermal gas. The morphology becomes more flattened or squashed when the effect of CRs is included and the difference is a result of CR diffusion as can also be seen from Fig.\ref{fig:diffusion}. As a result, in TH runs, the contact discontinuity (CD) remains near the free wind region and the wind created by the subsequent star formation episodes hits the CD and it breaks into high density gas clumps. However, due to the flattened nature of the free wind region in CR runs, the CD cannot be hit directly by the wind produced later. Taken all together, it is evident that the CR diffusion inhibits the clump formation either by washing out the small perturbations or by changing the morphology of the outflow. We therefore suggest that suppression of cloud formation due to CRs can reduce the SFR.

\section{Discussion and Conclusions}\label{sec:discussion}
With the help of our results from idealized simulations of galaxies of three different halo masses and three different SFRs we have shown that there is no significant effect of CRs in the {\it dynamics} of galactic outflow, but it alters the temperature distribution of gas. Here we explore the impact of the parameters we have used and compare the results with previous works.

\subsection{Comparison with previous works}\label{sec:compare_prev}

\begin{footnotesize}
\begin{table*}
\begin{center}
\caption{\large{Comparison of a few recent studies}}\label{tab:comparison}
\begin{tabular}{cccccc}
\hline
\hline
\makecell{Some recent\\results} & \makecell{Simulation\\ time ($t_{\rm sim}$)} & \makecell{Halo properties} & \makecell{SFR} & \makecell{Mass loading factor ($\eta$) with CR} & \makecell{Effect on SFR by CR} \\
\hline
\\
\makecell{Booth \\et al.\\ (2013)} & 500 Myr & \makecell{$M_{\rm halo} = 10^{12}$ $\& 10^9 M_{\odot}$\\$T_{\rm CGM} =$ NM$^\star$ \\ $n_{\rm CGM} =$ NM} & discrete$^{\star \star}$ & \makecell{For MW, $\eta \sim 0.5$ for both\\ feedback model. For SMC, $\eta$ is \\10 times higher for CR feedback} & \makecell{For both MW \& SMC\\ SFR is more suppressed \\by CR feedback} \\
\\
\makecell{Fujita \& \\ Mac \\ Low (2018)} & 2.2 Myr & \makecell{$M_{\rm halo} = 5\times$ $10^{12} M_{\odot}$\\$T_{\rm CGM} =$ NM \\ $n_{\rm CGM} =$ NM} & continuous & \makecell{CR pressure initially loads\\ about twice as much mass\\ as thermal pressure alone at\\ blowout, but the mass loading \\becomes similar afterwards} & \makecell{feedback is not\\coupled to SFR,\\no effect \\on SFR} \\
\\
\makecell{Butsky \& \\ Quinn (2018)} & 13 Gyr & \makecell{$M_{\rm halo} = 10^{12}$ $M_{\odot}$\\$T_{\rm CGM} =$ NM \\ $n_{\rm CGM} =$ NM} &  \makecell{discrete} & \makecell{CR pressure support lifts\\ thermal gas higher out of\\ the gravitational potential\\ well; no comment on $\eta$} & \makecell{higher value of\\ $p_{\rm cr}/p_{\rm th}$ implies more\\ suppression of SFR\\ by CR feedback }\\
\\
\makecell {Samui \\et al.\\ (2018)} & 10 Gyr & \makecell{$M_{\rm halo} = 10^{8--11}$ $M_{\odot}$\\$T_{\rm CGM} = T_{\rm vir}$ \\ $n_{\rm CGM} =$ follows a\\ beta model} & analytical & \makecell{$\eta$ = 33 for M8\\ \& $\eta =$ 0.7 for M11 \\ (assumed)} & \makecell{uses a SF model\\ dependent on\\ assumed values of $\eta$}\\
\\
\makecell{Jacob \\et al.\\(2018)} & 6 Gyr & \makecell{$M_{\rm halo} = 10^{10--13}$ $M_{\odot}$\\$T_{\rm CGM} =$ NM \\ $n_{\rm CGM} =$ NM}   & discrete & \makecell{$\eta \propto M^{-1}_{\rm vir} - M^{-2}_{\rm vir}$} & \makecell{no comparison \\with thermal}\\
\\
\makecell{Dashyan \& \\ Dubois (2020)} & 250 Myr & \makecell{$M_{\rm halo} = 10^{10} \& 10^{11}$ $M_{\odot}$\\$T_{\rm CGM} = 10^6$ K \\ $n_{\rm CGM} = 10^{-6}$ cm$^{-3}$}  & \makecell{discrete} & \makecell{Iso diffusion of CR\\ increases mass\\ loading} & \makecell{reduced by a\\ factor of 2}\\
\\
\makecell{Hopkins \\et al.\\(2020)} & 14 Gyr & \makecell{$M_{\rm halo} = 10^{9--12}$ $M_{\odot}$\\$T_{\rm CGM} =$ NM \\ $n_{\rm CGM} =$ NM}  & \makecell{discrete} & \makecell{volume filling factor\\ of outflow increases,\\ however, the rate\\ of outflow is similar\\ near the disk} & \makecell{suppressed 4\\ times than\\ thermal where CR\\ pressure dominate}\\
\\
Our Model & \makecell{50 Myr,\\210Myr (for\\fiducial case)} & \makecell{$M_{\rm halo} = 10^{8}, 10^{11}$, $10^{12} M_{\odot}$\\$T_{\rm CGM} = T_{\rm vir}$ \\ $n_{\rm CGM} =10^{-3}-10^{-4}$ cm$^{-3}$}  & continuous & similar as thermal & \makecell{feedback is not\\coupled to SFR,\\no effect on SFR}\\
\\
\hline
\hline
\end{tabular}
\end{center}
\hspace{-14.5cm} $\star$ NM = Not Mentioned\\
\hspace{-14.2cm} $\star \star$ For details see section \ref{sec:SFR_choice}
\end{table*}
\end{footnotesize}

We find that there is no significant difference in the time evolution  of mass loading factor between runs with and without CR in all cases. This is in agreement with \citet{Fujita2018}, who considered a galaxy of mass $5 \times 10^{12}$ $M_\odot$ (although their mass outflow rate includes all mass moving upward with $v_{\rm z,min} = 10$ km s$^{-1}$). However our result is in partial agreement with \citet{Booth2013}, who did not find any significant difference in mass loading factor after the inclusion of CR in Milky Way-sized galaxy, but found a ten-fold increase  for a SMC-type (with mass $2 \times 10^9$ $M_\odot$) galaxy. It is, however, not easy to trace the reason for the difference, because the halo gas temperature and density used by them are not mentioned.

One result that is similar to \citet{Booth2013}, \citet{Salem2014} and \citet{Butsky2018} is that we find the multiphase structure of the outflow in both the TH and TH+CR runs, however the amount of very cold gas ($T < 10^4$ K) is more in the models where CRs are included. We have traced the reason of this phenomenon to the reduction of gas temperature in the case of CR, which enhances cooling rate for $T < 10^7$ K.

Our results, however, contradict the analytical results by \citet{Samui2018}. They suggested that galactic winds driven by CRs travel a larger distance compared to thermally driven winds in massive galaxies. They reported that in a galaxy of mass $\sim 10^8 M_\odot$, a steady outflow is possible only with the inclusion of CR since 
they lose less energy due to adiabatic expansion because of the softer equation of state. However, even in the analytical calculation of blastwaves in the presence of CRs by \citet{Chevalier1983}, it was shown that shocks propagate to shorter distances compared to the case of purely thermal gas. Our results are consistent with these theoretical expectations. In Table \ref{tab:comparison} we summarise the differences among some previous works.

\subsection{Dependence on the choice of SFR}\label{sec:SFR_choice}
There is a basic difference in the way star formation is implemented in our work and most of the previous works. We assumed that star forms \textit{continuously} in the galactic disk, for the first few Myr with a time-independent SFR, whereas others assume that stars form \textit{discretely} whenever a grid point in their simulation reaches a critical density. Such a grid point is then called a `star particle'. Each star particle represents an ensemble of stars which follow an initial mass function (IMF). Depending on the IMF, in a given time duration $\Delta t$, there will be a number of SNe. For each SNe, they inject $10^{51}$ erg of energy and thereafter, the density of the grid point is updated accordingly. It can be understood that for the same IMF, a continuous SFR will inject more momentum into the ISM compared to a method where star formation happens discretely in some cells.

In the case of discrete star formation in dense grid points, SFR at a given time is dependent on the SFR at earlier time, which means there is a feedback channel which couples with the SFR. All previous works except \citet{Fujita2018} used discrete star formation with feedback and found that outflows with CRs are more effective for increasing mass loading and suppressing SFR compared to purely thermal outflow. On the other hand we find that CRs do not have any significant effect on the dynamics of the outflowing gas and our result matches with \citet{Fujita2018}. This contrast  suggests that the result is dependent on the choice of how SFR is implemented in the setup which lead us to the discussion in section \ref{sec:cold_clumps}.

\subsection{Choice of CR injection region}\label{sec:diff_injection}
\begin{figure}
\centering
\includegraphics[width=0.52\textwidth]{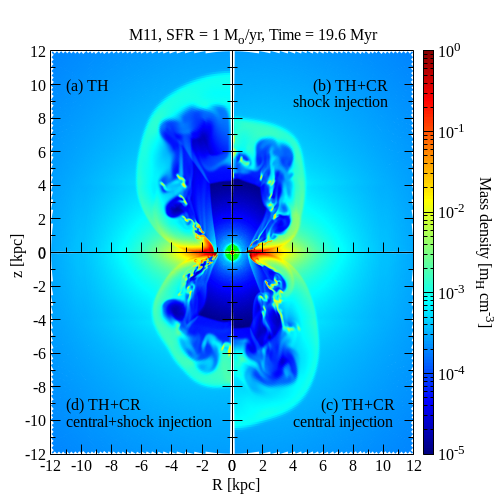}
\caption{The 2-D density plot for the fiducial case (M11, SFR = 1 $\sfr$) at $\approx 20$ Myr are shown for different CR injection schemes. In the fiducial TH+CR case, CRs are injected as pressure at the shock fronts (shock injection). CRs can also be injected at the centre of the thermal injection region (central injection) or as a combination of both shock and central injection. We see that the dynamical effect of the CRs is to shrink the outer shock in a more pronounced manner when CRs are included at the shock fronts.}
\label{fig:injection}
\end{figure}
\begin{figure}
\centering
\includegraphics[width=0.48\textwidth]{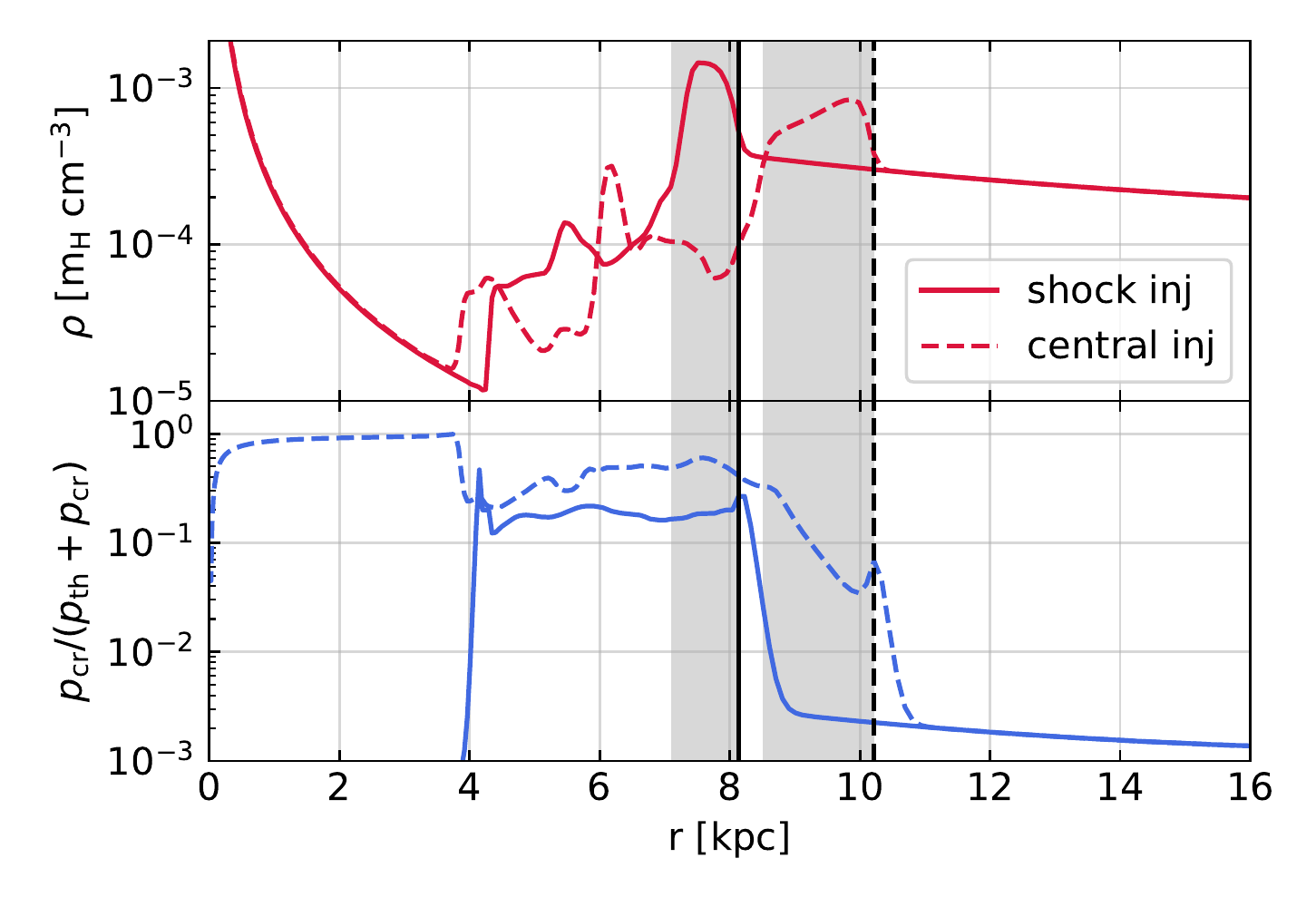}
\caption{For the fiducial case (M11 galaxy, SFR=1 $\sfr$, along $\theta = 20^{\circ}$), density (red curves) and CR pressure fraction ($p_{\rm cr}/(p_{\rm th}+p_{\rm cr})$, blue curves) are shown at $\approx 20$ Myr in case of shock injection (solid curves) and central injection (dashed curves). The outer shock position in these two cases are marked with two black lines (dashed: central injection, solid: shock injection) and the shocked gas regions are shown by grey shaded area. The density jump at the outer shock is higher in case of shock injection. We also note that $p_{\rm cr}/(p_{\rm th}+p_{\rm cr})$ is less near the outer shock position in the central injection case.} \label{fig:1d_diffCRinj}
\end{figure}

In this study, we have injected CRs at the locations of shocks, which differs from most previous studies where a fraction of the total energy is directly injected into CRs at the same location where SNe energy is injected (see e.g., \citealt{Booth2013,Salem2014,Butsky2018,Fujita2018}). We find that these two injection methods can affect the morphology of outflowing gas, which is shown in Fig. \ref{fig:injection}. Note that, to distinguish these two methods, we have labelled them by {\it shock injection} (as described in section \ref{sec:injection}) and {\it central injection} (where $20\%$ of total input energy is injected into CRs within $r_{\rm inj}$). In addition to these two methods, we also show a combined case - central + shock injections.

Fig. \ref{fig:injection} shows that outer shock travels a shorter distance in the shock injection case (panel b) compared to the central injection case (panel c). For both cases, the reverse shock is located at a similar location (along  $z$ axis, it is located at $\approx 4.5$ kpc).  However, the gas swept-up by outer shock is denser in the shock injection model compared to the central injection model. The reason behind this can be understood as follows.
\begin{itemize}
\item In case of shock injection, the CRs are continuously injected at the location of shocks, thereby it maintains a constant CR pressure fraction in the shocked gas region. Whereas in the central injection case, CR energy in the shocked gas evolves with time, and we find that the CR energy is quite high near $r_{\rm inj}$ and also in the shocked wind region, however, it drops gradually near the outer shock (thereby it reduces CR pressure gradient w.r.t. the CGM). It can be seen by comparing solid and dashed blue curves (representing shock and central injections respectively) in Fig. \ref{fig:1d_diffCRinj}, which represent the radial distribution of CR pressure fraction at $\theta = 20 \degree$. Therefore, the leakage of CR energy from the outer shocked gas to the halo is more effective in the shock injection case compared to the central injection case.
\item Due to efficient CR diffusion in the shock injection case, the shocked gas of the outflow gets compressed to a higher density than the central injection case (compare solid and dashed red curves in Fig. \ref{fig:1d_diffCRinj}). This effect also enhances the radiative cooling of the gas, swept-up by outer shock, since the cooling rate of the gas $\propto \rho^{2}$. As a result, in the shock injection case, the outer shock travels a shoter distance.
\end{itemize}
Therefore, both CR diffusion and radiative cooling of the gas play a crucial role in determining the location of outer shock.

\subsection{Dependence on CR diffusion coefficient}\label{sec:diffusion}
\begin{figure*}
\includegraphics[width=0.6\textwidth]{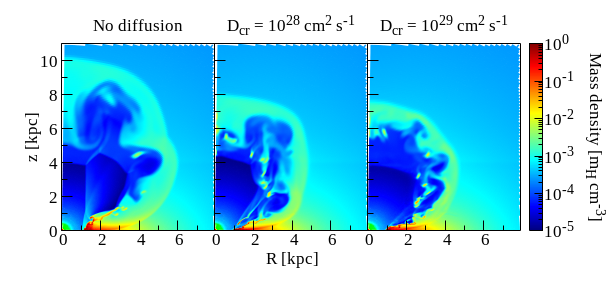}%
\includegraphics[width=0.4\textwidth]{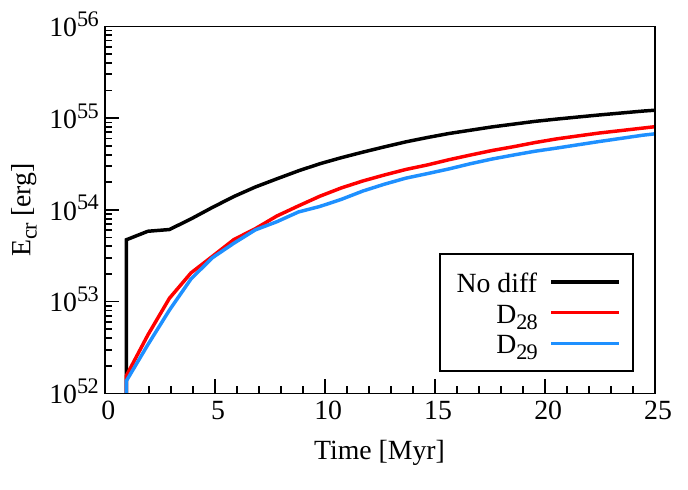}
\caption{Position of outer shock in the 2D density plots at $\approx 20$ Myr for M11 galaxy, SFR = 1 $\sfr$ including CRs but with different values of CR diffusion coefficients ($D_{\rm cr}$). It can be seen that the outer shock traverses a shorter distance with increasing values of $D_{\rm cr}$. In a separate panel (right side) we also show how CR energy ($E_{\rm cr}$) within outer shock changes depending upon the value of $D_{\rm cr}$.
}
\label{fig:diffusion}
\end{figure*}

To understand the importance of CR diffusion, we show  2-D density plots for M11 galaxy, SFR = 1 $\sfr$, at $\approx 20$ Myr, for three different values of diffusion coefficient ($D_{\rm cr}$) in Fig. \ref{fig:diffusion}. We note that the position of the outer shock decreases with increasing $D_{\rm cr}$. This confirms our explanation that the energy contained within the bubble leaks more with increasing CR diffusion as described in Section \ref{sec:fwd_shock}. This can also be verified from Fig. \ref{fig:diffusion}, right panel, where it shows that the CR energy within the bubble decreases for higher value of diffusion coefficient ($D_{\rm cr}$).

\subsection{Limitations of our work}\label{sec:limitations}
\begin{itemize}
\item A 3D simulation is more favourable to study the morphology of the outflowing gas, however, it is computationally expensive and a 2D version is unlikely to change the global dynamical features. We would like to point out that \citet{Fujita2018} found that the kinematic behaviour of the outflow in their 3D simulation agrees well with the 2D version \citep{Fujita2009}. Hence it is expected that our results will not change in a more realistic 3D simulation.
\item The results described here are valid for our assumption of CR transport mechanism i.e. isotropic diffusion which is a necessary assumption in CR-hydro simulations. When the magnetic field is considered, the effect of CR diffusion is mostly dominated by the choice of the diffusion coefficient parallel to the magnetic field ($k_{\rm \parallel}$) since the diffusion coefficient perpendicular to the magnetic field is much smaller than $k_{\rm \parallel}$. However, given the uncertainty in the value of parallel/perpendicular CR diffusion coefficient, the choice of isotropic diffusion is still a good assumption.
\end{itemize}

\section{Summary}\label{sec:summary}
We have performed idealized hydrodynamical (HD) simulation of three galaxies with virial mass $10^{8}$, $10^{11}$ and $10^{12}$ $M_\odot$ for constant star formation rates $0.1$, $1$ and $10$ $\sfr$. Our aim has been to find the dynamical effects of CRs on galactic scale outflow of gas in the early evolutionary stage ($\le 50$ Myr). Hence we have performed two sets of simulations for each case, one, in which thermal pressure alone drives the outflow (TH runs) and another, in which $20\%$ of the total pressure at the shocks is attributed to CRs (TH + CR runs). 
Comparing these two sets of simulations, our results are summarized as follows:
\begin{enumerate}
\item We find that the outflow reaches a shorter distance when CRs are included in the model compared to purely thermally driven outflows. The main reason behind this is the diffusion of CRs out of the bubble (within outer shock, where CR pressure is high, compared to the halo gas outside). We also note that the inclusion of CRs decrease the value of the effective adiabatic index of the combined fluid from $5/3$ and from the energy conservation it can be shown that the volume of the outflowing gas is less if it is driven by CR pressure instead of thermal pressure. This result is more prominent for higher mass galaxies and higher SFRs. However, for M8 galaxy, we do not see any difference in the outer shock position between the runs with and without CRs. This is because the speed of outflow in this case is few hundred km s$^{-1}$, one order of magnitude larger than the escape speed ($\sim 10$ km/s) of M8 galaxy. The outflow moves so fast that the effect of diffusion is not observed in this case.
 
\item The mass loading factor (ratio of the mass outflow rate to SFR) has been calculated in two ways, one in which the mass outflow rate is estimated through a plane parallel to the galactic disk (above and below) and another in which it has been calculated through a spherical surface. In both cases, we do not find any significant difference between the mass loading factors in TH and TH+CR cases. This result is valid for all the galaxies for all constant SFRs. However, for higher mass galaxies and lower SFR there is a trend to show a few negative values in the time evolution of the mass loading factor owing to the strong gravitational potential of  high mass galaxies.

\item We have studied the distribution of gas in different temperature bins, as well as distribution of gas mass in different density and temperature bins within the simulation box.  We find that the amount of cold gas ($\sim 10^4$ K) is more for models which includes CR. This arises from the fact that a part of the thermal energy is attributed to CR energy in TH+CR models, hence the effective temperature of the outflowing gas decreases and increases its radiative cooling rate.

\item We have investigated the long term evolution of the outflow by performing a simulation of the fiducial case (M11 galaxy) for a duartion of 210 Myr with periodic star formation events. After $\approx$ 160 Myr, we observe that a few high density and low temperature gas clumps form in TH run, which are abesnt in TH+CR. 
We have identified a possible explanation for the suppression of these clump formation by the diffusive nature of CRs. Our discussion suggets that the formation of these clumps and their infall on the galactic disk can decrease the efficiency of feedback (by enhancing star formation due to the presence of cold clumps) which is reported by previous results. 
   
\end{enumerate}

In a nutshell, we found that CRs do not change the dynamics of the outflowing gas significantly in the early stage of galactic outflow ($\le 50$ Myr), however the temperature of the outflowing gas decreases in the presence of CRs. Moreover, the idealized simulation provided us a suitable platform to study the reason behind the higher efficiency of CR driven feedback as reported in earlier works.
\section*{Acknowledgements}
We would like to thank Kartick C. Sarkar and Prateek Sharma for valuable discussions, and the Supercomputing Education and Research Centre (SERC), IISc, where a part of the simulations was performed by Cray XC40-SahasraT cluster.
\section*{Data availability}
No new data were generated or analysed in support of this research.

\begin{footnotesize}
\bibliographystyle{mnras}
\nocite{}
\bibliography{reference}

\begin{thebibliography}{}
\makeatletter
\relax
\def\mn@urlcharsother{\let\do\@makeother \do\$\do\&\do\#\do\^\do\_\do\%\do\~}
\def\mn@doi{\begingroup\mn@urlcharsother \@ifnextchar [ {\mn@doi@}
  {\mn@doi@[]}}
\def\mn@doi@[#1]#2{\def\@tempa{#1}\ifx\@tempa\@empty \href
  {http://dx.doi.org/#2} {doi:#2}\else \href {http://dx.doi.org/#2} {#1}\fi
  \endgroup}
\def\mn@eprint#1#2{\mn@eprint@#1:#2::\@nil}
\def\mn@eprint@arXiv#1{\href {http://arxiv.org/abs/#1} {{\tt arXiv:#1}}}
\def\mn@eprint@dblp#1{\href {http://dblp.uni-trier.de/rec/bibtex/#1.xml}
  {dblp:#1}}
\def\mn@eprint@#1:#2:#3:#4\@nil{\def\@tempa {#1}\def\@tempb {#2}\def\@tempc
  {#3}\ifx \@tempc \@empty \let \@tempc \@tempb \let \@tempb \@tempa \fi \ifx
  \@tempb \@empty \def\@tempb {arXiv}\fi \@ifundefined
  {mn@eprint@\@tempb}{\@tempb:\@tempc}{\expandafter \expandafter \csname
  mn@eprint@\@tempb\endcsname \expandafter{\@tempc}}}

\bibitem[\protect\citeauthoryear{{Agertz}, {Kravtsov}, {Leitner}  \&
  {Gnedin}}{{Agertz} et~al.}{2013}]{Agertz2013}
{Agertz} O.,  {Kravtsov} A.~V.,  {Leitner} S.~N.,   {Gnedin} N.~Y.,  2013,
  \mn@doi [\apj] {10.1088/0004-637X/770/1/25}, \href
  {https://ui.adsabs.harvard.edu/abs/2013ApJ...770...25A} {770, 25}

\bibitem[\protect\citeauthoryear{{Booth}, {Agertz}, {Kravtsov}  \&
  {Gnedin}}{{Booth} et~al.}{2013}]{Booth2013}
{Booth} C.~M.,  {Agertz} O.,  {Kravtsov} A.~V.,   {Gnedin} N.~Y.,  2013,
  \mn@doi [\apj] {10.1088/2041-8205/777/1/L16}, \href
  {https://ui.adsabs.harvard.edu/abs/2013ApJ...777L..16B} {777, L16}

\bibitem[\protect\citeauthoryear{{Breitschwerdt}, {McKenzie}  \&
  {Voelk}}{{Breitschwerdt} et~al.}{1991}]{Breitschwerdt1991}
{Breitschwerdt} D.,  {McKenzie} J.~F.,   {Voelk} H.~J.,  1991, \aap, \href
  {https://ui.adsabs.harvard.edu/abs/1991A%26A...245...79B} {245, 79}

\bibitem[\protect\citeauthoryear{{Butsky} \& {Quinn}}{{Butsky} \&
  {Quinn}}{2018}]{Butsky2018}
{Butsky} I.~S.,  {Quinn} T.~R.,  2018, \mn@doi [\apj]
  {10.3847/1538-4357/aaeac2}, \href
  {https://ui.adsabs.harvard.edu/abs/2018ApJ...868..108B} {868, 108}

\bibitem[\protect\citeauthoryear{{Chevalier}}{{Chevalier}}{1983}]{Chevalier1983}
{Chevalier} R.~A.,  1983, \mn@doi [\apj] {10.1086/161338}, \href
  {https://ui.adsabs.harvard.edu/abs/1983ApJ...272..765C} {272, 765}

\bibitem[\protect\citeauthoryear{{Dashyan} \& {Dubois}}{{Dashyan} \&
  {Dubois}}{2020}]{Dashyan2020}
{Dashyan} G.,  {Dubois} Y.,  2020, arXiv e-prints, \href
  {https://ui.adsabs.harvard.edu/abs/2020arXiv200309900D} {p. arXiv:2003.09900}

\bibitem[\protect\citeauthoryear{{Drury} \& {Voelk}}{{Drury} \&
  {Voelk}}{1981}]{Drury1981}
{Drury} L.~O.,  {Voelk} J.~H.,  1981, \mn@doi [\apj] {10.1086/159159}, \href
  {https://ui.adsabs.harvard.edu/abs/1981ApJ...248..344D} {248, 344}

\bibitem[\protect\citeauthoryear{{Efstathiou}}{{Efstathiou}}{2000}]{Efstathiou2000}
{Efstathiou} G.,  2000, \mn@doi [\mnras] {10.1046/j.1365-8711.2000.03665.x},
  \href {https://ui.adsabs.harvard.edu/abs/2000MNRAS.317..697E} {317, 697}

\bibitem[\protect\citeauthoryear{{Everett}, {Zweibel}, {Benjamin}, {McCammon},
  {Rocks}  \& {Gallagher}}{{Everett} et~al.}{2008}]{Everett2008}
{Everett} J.~E.,  {Zweibel} E.~G.,  {Benjamin} R.~A.,  {McCammon} D.,  {Rocks}
  L.,   {Gallagher} III J.~S.,  2008, \mn@doi [\apj] {10.1086/524766}, \href
  {https://ui.adsabs.harvard.edu/abs/2008ApJ...674..258E} {674, 258}

\bibitem[\protect\citeauthoryear{{Fujita} \& {Mac Low}}{{Fujita} \& {Mac
  Low}}{2018}]{Fujita2018}
{Fujita} A.,  {Mac Low} M.-M.,  2018, \mn@doi [\mnras] {10.1093/mnras/sty715},
  \href {https://ui.adsabs.harvard.edu/abs/2018MNRAS.477..531F} {477, 531}

\bibitem[\protect\citeauthoryear{{Fujita}, {Martin}, {Mac Low}, {New}  \&
  {Weaver}}{{Fujita} et~al.}{2009}]{Fujita2009}
{Fujita} A.,  {Martin} C.~L.,  {Mac Low} M.-M.,  {New} K. C.~B.,   {Weaver} R.,
   2009, \mn@doi [\apj] {10.1088/0004-637X/698/1/693}, \href
  {https://ui.adsabs.harvard.edu/abs/2009ApJ...698..693F} {698, 693}

\bibitem[\protect\citeauthoryear{{Girichidis} et~al.,}{{Girichidis}
  et~al.}{2016}]{Girichidis2016a}
{Girichidis} P.,  et~al., 2016, \mn@doi [\mnras] {10.1093/mnras/stv2742}, \href
  {https://ui.adsabs.harvard.edu/abs/2016MNRAS.456.3432G} {456, 3432}

\bibitem[\protect\citeauthoryear{Guo \& Oh}{Guo \& Oh}{2008}]{Guo2008}
Guo F.,  Oh S.~P.,  2008, \mn@doi [\mnras] {10.1111/j.1365-2966.2007.12692.x},
  384, 251

\bibitem[\protect\citeauthoryear{{Gupta}, {Nath}, {Sharma}  \&
  {Eichler}}{{Gupta} et~al.}{2018a}]{Gupta2018Jan}
{Gupta} S.,  {Nath} B.~B.,  {Sharma} P.,   {Eichler} D.,  2018a, \mn@doi
  [\mnras] {10.1093/mnras/stx2427}, \href
  {https://ui.adsabs.harvard.edu/abs/2018MNRAS.473.1537G} {473, 1537}

\bibitem[\protect\citeauthoryear{{Gupta}, {Nath}  \& {Sharma}}{{Gupta}
  et~al.}{2018b}]{Gupta2018Oct}
{Gupta} S.,  {Nath} B.~B.,   {Sharma} P.,  2018b, \mn@doi [\mnras]
  {10.1093/mnras/sty1846}, \href
  {https://ui.adsabs.harvard.edu/abs/2018MNRAS.479.5220G} {479, 5220}

\bibitem[\protect\citeauthoryear{{Gupta}, {Sharma}  \& {Mignone}}{{Gupta}
  et~al.}{2019}]{Gupta2019}
{Gupta} S.,  {Sharma} P.,   {Mignone} A.,  2019, arXiv e-prints, \href
  {https://ui.adsabs.harvard.edu/abs/2019arXiv190607200G} {p. arXiv:1906.07200}

\bibitem[\protect\citeauthoryear{{Hanasz} \& {Lesch}}{{Hanasz} \&
  {Lesch}}{2003}]{Hanasz2003}
{Hanasz} M.,  {Lesch} H.,  2003, \mn@doi [\aap] {10.1051/0004-6361:20031433},
  \href {https://ui.adsabs.harvard.edu/abs/2003A&A...412..331H} {412, 331}

\bibitem[\protect\citeauthoryear{{Heckman}, {Alexandroff}, {Borthakur},
  {Overzier}  \& {Leitherer}}{{Heckman} et~al.}{2015}]{Heckman2015}
{Heckman} T.~M.,  {Alexandroff} R.~M.,  {Borthakur} S.,  {Overzier} R.,
  {Leitherer} C.,  2015, \mn@doi [\apj] {10.1088/0004-637X/809/2/147}, \href
  {https://ui.adsabs.harvard.edu/abs/2015ApJ...809..147H} {809, 147}

\bibitem[\protect\citeauthoryear{{Hopkins}, {Quataert}  \& {Murray}}{{Hopkins}
  et~al.}{2012}]{Hopkins2012}
{Hopkins} P.~F.,  {Quataert} E.,   {Murray} N.,  2012, \mn@doi [\mnras]
  {10.1111/j.1365-2966.2012.20593.x}, \href
  {https://ui.adsabs.harvard.edu/abs/2012MNRAS.421.3522H} {421, 3522}

\bibitem[\protect\citeauthoryear{{Ipavich}}{{Ipavich}}{1975}]{Ipavich1975}
{Ipavich} F.~M.,  1975, \mn@doi [\apj] {10.1086/153397}, \href
  {https://ui.adsabs.harvard.edu/abs/1975ApJ...196..107I} {196, 107}

\bibitem[\protect\citeauthoryear{{Jacob}, {Pakmor}, {Simpson}, {Springel}  \&
  {Pfrommer}}{{Jacob} et~al.}{2018}]{Jacob2018}
{Jacob} S.,  {Pakmor} R.,  {Simpson} C.~M.,  {Springel} V.,   {Pfrommer} C.,
  2018, \mn@doi [\mnras] {10.1093/mnras/stx3221}, \href
  {https://ui.adsabs.harvard.edu/abs/2018MNRAS.475..570J} {475, 570}

\bibitem[\protect\citeauthoryear{{Krumholz} \& {Thompson}}{{Krumholz} \&
  {Thompson}}{2012}]{Krumholz2012}
{Krumholz} M.~R.,  {Thompson} T.~A.,  2012, \mn@doi [\apj]
  {10.1088/0004-637X/760/2/155}, \href
  {https://ui.adsabs.harvard.edu/abs/2012ApJ...760..155K} {760, 155}

\bibitem[\protect\citeauthoryear{{Kulsrud} \& {Pearce}}{{Kulsrud} \&
  {Pearce}}{1969}]{Kulsrud1969}
{Kulsrud} R.,  {Pearce} W.~P.,  1969, \mn@doi [\apj] {10.1086/149981}, \href
  {https://ui.adsabs.harvard.edu/abs/1969ApJ...156..445K} {156, 445}

\bibitem[\protect\citeauthoryear{{Martizzi}, {Fielding}, {Faucher-Gigu{\`e}re}
  \& {Quataert}}{{Martizzi} et~al.}{2016}]{Martizzi2016}
{Martizzi} D.,  {Fielding} D.,  {Faucher-Gigu{\`e}re} C.-A.,   {Quataert} E.,
  2016, \mn@doi [\mnras] {10.1093/mnras/stw745}, \href
  {https://ui.adsabs.harvard.edu/abs/2016MNRAS.459.2311M} {459, 2311}

\bibitem[\protect\citeauthoryear{{McCourt}, {Oh}, {O'Leary}  \&
  {Madigan}}{{McCourt} et~al.}{2018}]{McCourt2018}
{McCourt} M.,  {Oh} S.~P.,  {O'Leary} R.,   {Madigan} A.-M.,  2018, \mn@doi
  [\mnras] {10.1093/mnras/stx2687}, \href
  {https://ui.adsabs.harvard.edu/abs/2018MNRAS.473.5407M} {473, 5407}

\bibitem[\protect\citeauthoryear{{Mignone}, {Bodo}, {Massaglia}, {Matsakos},
  {Tesileanu}, {Zanni}  \& {Ferrari}}{{Mignone} et~al.}{2007}]{Mignone2007}
{Mignone} A.,  {Bodo} G.,  {Massaglia} S.,  {Matsakos} T.,  {Tesileanu} O.,
  {Zanni} C.,   {Ferrari} A.,  2007, \mn@doi [\apjs] {10.1086/513316}, \href
  {https://ui.adsabs.harvard.edu/abs/2007ApJS..170..228M} {170, 228}

\bibitem[\protect\citeauthoryear{{Miyamoto} \& {Nagai}}{{Miyamoto} \&
  {Nagai}}{1975}]{Miyamoto1975}
{Miyamoto} M.,  {Nagai} R.,  1975, \pasj, \href
  {https://ui.adsabs.harvard.edu/abs/1975PASJ...27..533M} {27, 533}

\bibitem[\protect\citeauthoryear{{Murray}, {Quataert}  \& {Thompson}}{{Murray}
  et~al.}{2005}]{Murray2005}
{Murray} N.,  {Quataert} E.,   {Thompson} T.~A.,  2005, \mn@doi [\apj]
  {10.1086/426067}, \href
  {https://ui.adsabs.harvard.edu/abs/2005ApJ...618..569M} {618, 569}

\bibitem[\protect\citeauthoryear{{Navarro}, {Frenk}  \& {White}}{{Navarro}
  et~al.}{1997}]{Navarro1997}
{Navarro} J.~F.,  {Frenk} C.~S.,   {White} S. D.~M.,  1997, \mn@doi [\apj]
  {10.1086/304888}, \href
  {https://ui.adsabs.harvard.edu/abs/1997ApJ...490..493N} {490, 493}

\bibitem[\protect\citeauthoryear{{Pfrommer}, {Pakmor}, {Schaal}, {Simpson}  \&
  {Springel}}{{Pfrommer} et~al.}{2017}]{Pfrommer2017}
{Pfrommer} C.,  {Pakmor} R.,  {Schaal} K.,  {Simpson} C.~M.,   {Springel} V.,
  2017, \mn@doi [\mnras] {10.1093/mnras/stw2941}, \href
  {https://ui.adsabs.harvard.edu/abs/2017MNRAS.465.4500P} {465, 4500}

\bibitem[\protect\citeauthoryear{{Recchia}, {Blasi}  \& {Morlino}}{{Recchia}
  et~al.}{2016}]{Recchia2016}
{Recchia} S.,  {Blasi} P.,   {Morlino} G.,  2016, \mn@doi [\mnras]
  {10.1093/mnras/stw1966}, \href
  {https://ui.adsabs.harvard.edu/abs/2016MNRAS.462.4227R} {462, 4227}

\bibitem[\protect\citeauthoryear{{Rosdahl}, {Schaye}, {Teyssier}  \&
  {Agertz}}{{Rosdahl} et~al.}{2015}]{Rosdahl2015}
{Rosdahl} J.,  {Schaye} J.,  {Teyssier} R.,   {Agertz} O.,  2015, \mn@doi
  [\mnras] {10.1093/mnras/stv937}, \href
  {https://ui.adsabs.harvard.edu/abs/2015MNRAS.451...34R} {451, 34}

\bibitem[\protect\citeauthoryear{{Ruszkowski}, {Yang}  \&
  {Zweibel}}{{Ruszkowski} et~al.}{2017}]{Ruszkowski2017}
{Ruszkowski} M.,  {Yang} H. Y.~K.,   {Zweibel} E.,  2017, \mn@doi [\apj]
  {10.3847/1538-4357/834/2/208}, \href
  {https://ui.adsabs.harvard.edu/abs/2017ApJ...834..208R} {834, 208}

\bibitem[\protect\citeauthoryear{{Salem} \& {Bryan}}{{Salem} \&
  {Bryan}}{2014}]{Salem2014}
{Salem} M.,  {Bryan} G.~L.,  2014, \mn@doi [\mnras] {10.1093/mnras/stt2121},
  \href {https://ui.adsabs.harvard.edu/abs/2014MNRAS.437.3312S} {437, 3312}

\bibitem[\protect\citeauthoryear{{Salpeter}}{{Salpeter}}{1955}]{Salpeter1955}
{Salpeter} E.~E.,  1955, \mn@doi [\apj] {10.1086/145971}, \href
  {https://ui.adsabs.harvard.edu/abs/1955ApJ...121..161S} {121, 161}

\bibitem[\protect\citeauthoryear{{Samui}, {Subramanian}  \& {Srianand
  }}{{Samui} et~al.}{2010}]{Samui2010}
{Samui} S.,  {Subramanian} K.,   {Srianand } R.,  2010, \mn@doi [\mnras]
  {10.1111/j.1365-2966.2009.16099.x}, \href
  {https://ui.adsabs.harvard.edu/abs/2010MNRAS.402.2778S} {402, 2778}

\bibitem[\protect\citeauthoryear{{Samui}, {Subramanian}  \& {Srianand
  }}{{Samui} et~al.}{2018}]{Samui2018}
{Samui} S.,  {Subramanian} K.,   {Srianand } R.,  2018, \mn@doi [\mnras]
  {10.1093/mnras/sty287}, \href
  {https://ui.adsabs.harvard.edu/abs/2018MNRAS.476.1680S} {476, 1680}

\bibitem[\protect\citeauthoryear{{Sarkar}, {Nath}, {Sharma}  \&
  {Shchekinov}}{{Sarkar} et~al.}{2015}]{Sarkar2015}
{Sarkar} K.~C.,  {Nath} B.~B.,  {Sharma} P.,   {Shchekinov} Y.,  2015, \mn@doi
  [\mnras] {10.1093/mnras/stu2760}, \href
  {https://ui.adsabs.harvard.edu/abs/2015MNRAS.448..328S} {448, 328}

\bibitem[\protect\citeauthoryear{{Sharma} \& {Nath}}{{Sharma} \&
  {Nath}}{2012}]{Sharma2012}
{Sharma} M.,  {Nath} B.~B.,  2012, \mn@doi [\apj] {10.1088/0004-637X/750/1/55},
  \href {https://ui.adsabs.harvard.edu/abs/2012ApJ...750...55S} {750, 55}

\bibitem[\protect\citeauthoryear{{Sharma}, {Roy}, {Nath}  \&
  {Shchekinov}}{{Sharma} et~al.}{2014}]{Sharma2014}
{Sharma} P.,  {Roy} A.,  {Nath} B.~B.,   {Shchekinov} Y.,  2014, \mn@doi
  [\mnras] {10.1093/mnras/stu1307}, \href
  {https://ui.adsabs.harvard.edu/abs/2014MNRAS.443.3463S} {443, 3463}

\bibitem[\protect\citeauthoryear{{Skinner} \& {Ostriker}}{{Skinner} \&
  {Ostriker}}{2015}]{Skinner2015}
{Skinner} M.~A.,  {Ostriker} E.~C.,  2015, \mn@doi [\apj]
  {10.1088/0004-637X/809/2/187}, \href
  {https://ui.adsabs.harvard.edu/abs/2015ApJ...809..187S} {809, 187}

\bibitem[\protect\citeauthoryear{{Strickland} \& {Heckman}}{{Strickland} \&
  {Heckman}}{2007}]{Strickland2007}
{Strickland} D.~K.,  {Heckman} T.~M.,  2007, \mn@doi [\apj] {10.1086/511174},
  \href {https://ui.adsabs.harvard.edu/abs/2007ApJ...658..258S} {658, 258}

\bibitem[\protect\citeauthoryear{{Sutherland} \& {Dopita}}{{Sutherland} \&
  {Dopita}}{1993}]{Sutherland1993}
{Sutherland} R.~S.,  {Dopita} M.~A.,  1993, \mn@doi [\apjs] {10.1086/191823},
  \href {https://ui.adsabs.harvard.edu/abs/1993ApJS...88..253S} {88, 253}

\bibitem[\protect\citeauthoryear{{Tang} \& {Wang}}{{Tang} \&
  {Wang}}{2005}]{Tang2005}
{Tang} S.,  {Wang} Q.~D.,  2005, \mn@doi [\apj] {10.1086/430875}, \href
  {https://ui.adsabs.harvard.edu/abs/2005ApJ...628..205T} {628, 205}

\bibitem[\protect\citeauthoryear{{Uhlig}, {Pfrommer}, {Sharma}, {Nath},
  {En{\ss}lin}  \& {Springel}}{{Uhlig} et~al.}{2012}]{Uhlig2012}
{Uhlig} M.,  {Pfrommer} C.,  {Sharma} M.,  {Nath} B.~B.,  {En{\ss}lin} T.~A.,
  {Springel} V.,  2012, \mn@doi [\mnras] {10.1111/j.1365-2966.2012.21045.x},
  \href {https://ui.adsabs.harvard.edu/abs/2012MNRAS.423.2374U} {423, 2374}

\bibitem[\protect\citeauthoryear{{Veilleux}, {Cecil}  \&
  {Bland-Hawthorn}}{{Veilleux} et~al.}{2005}]{Veilleux2005}
{Veilleux} S.,  {Cecil} G.,   {Bland-Hawthorn} J.,  2005, \mn@doi [\araa]
  {10.1146/annurev.astro.43.072103.150610}, \href
  {https://ui.adsabs.harvard.edu/abs/2005ARA&A..43..769V} {43, 769}

\bibitem[\protect\citeauthoryear{{Weaver}, {McCray}, {Castor}, {Shapiro}  \&
  {Moore}}{{Weaver} et~al.}{1977}]{Weaver1977}
{Weaver} R.,  {McCray} R.,  {Castor} J.,  {Shapiro} P.,   {Moore} R.,  1977,
  \mn@doi [\apj] {10.1086/155692}, \href
  {https://ui.adsabs.harvard.edu/abs/1977ApJ...218..377W} {218, 377}

\bibitem[\protect\citeauthoryear{{Zirakashvili}, {Breitschwerdt}, {Ptuskin}  \&
  {Voelk}}{{Zirakashvili} et~al.}{1996}]{Zirakashvili1996}
{Zirakashvili} V.~N.,  {Breitschwerdt} D.,  {Ptuskin} V.~S.,   {Voelk} H.~J.,
  1996, \aap, \href {https://ui.adsabs.harvard.edu/abs/1996A&A...311..113Z}
  {311, 113}

\makeatother
\end{thebibliography}
\end{footnotesize}

\appendix

\section{Cooling box}\label{app:coolbox}
\begin{figure}
\centering
\includegraphics[width=0.5\textwidth]{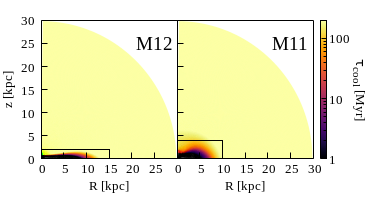}
\caption{Cooling timescale of M12 and M11 galaxy. The black outlined box shows the region where radiative cooling of the injected gas is only allowed, for details see Appendix \ref{app:coolbox}.}
\label{fig:cooltime}
\end{figure}
The cooling timescale of the gas in the galactic disk is usually shorter than the dynamical timescale that has been considered in this study. If there is no additional physical process to increase the energy of the gas then it would lose its thermal energy as time evolves. In reality, the disk gas is heated by photoionization radiation from stars and its temperature is maintained at $\sim 10^{4}$ K. In our simulations, since we do not include radiation, we keep the temperature of the disk at $\sim 10^{4}$ K for all runs. This is achieved by setting a box within which the cooling is allowed only for the the injected material (not the disk or halo gas in that region). 
In order to choose the width and height of this box, we first estimate the cooling time scale of the gas by using
\begin{eqnarray}
\tau_{\rm cool}& = & \frac{E_{\rm g}}{n_{\rm i}\, n_{\rm e}\, \Lambda_{\rm N}(T)} \nonumber\\
& \approx & 60 \,{\rm Myr}\, \left(\frac{T}{10^{6}{\rm K}}\right) \left( \frac{\rho}{10^{-3}\, m{\rm _H\, cm^{-3}}}\right)^{-1}\nonumber\\ && \left(\frac{\Lambda_{\rm N}}{10^{-22}\,{\rm erg\, cm^{3}\, s^{-1}}}\right)^{-1}\nonumber\\
\end{eqnarray}
where $\Lambda_{\rm N}$ is the normalized cooling rate, $T$ is the gas temperature, and $\rho$ is the gas density. Figure \ref{fig:cooltime} shows the cooling timescale for M12 and M11 galaxy and also the box within which the cooling of disk or halo gas is not allowed. The size of this box is chosen such that the cooling timescale of the gas beyond the box is longer than our simulation timescale. For the galaxy with mass $10^{8}\, M_{\rm \odot}$, the disk is expected to be small and hence we have not considered any box in this case.

\section{CR cooling}\label{app:CR_cooling}
In this study, we have not included CR energy loss. CRs can lose energy while interacting with matter. However, the cooling timescale of CRs due to these interactions is longer than the timescale used in our simulations. For example, the cooling timescale of CRs due to hadronic and leptonic interactions \citep{Guo2008} is given by
\begin{eqnarray}
\tau^{\rm cr}_{\rm cool} & = & \frac{e_{\rm cr}}{7.5 \times 10^{-16} n_{\rm H} e_{\rm cr}} \nonumber\\
& \approx & 4\times 10^4 \rm{Myr} \left(\frac{\rho}{10^{-3}\, m{\rm _H\, cm^{-3}}}\right)^{-1}
\end{eqnarray}
which is much longer than our simulation time scale.

\section{Convergence test}\label{app:high_resol}
\begin{figure}
\includegraphics[width=0.48\textwidth]{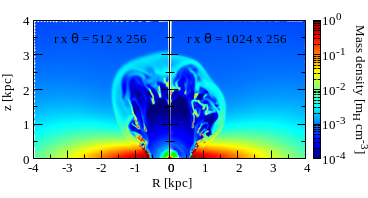}
\caption{The 2D density plots for M11 galaxy, SFR = 1 $\sfr$ at $\approx 7$ Myr show no significant difference in our fiducial resolution and a higher resolution (see text for details) considered.}
\label{fig:sch}
\end{figure}
The resolution used in all our simulations is described in Section \ref{sec:grid}. We performed a high resolution run for our fiducial case (M11 galaxy, SFR = 1 $\sfr$, at $\approx 7$ Myr). For this particular run, we use $1024$ grid points along $r$ direction and $256$ grid points along $\theta$ direction. We do not find any qualitative difference between the results of these two runs. This confirms that the resolution used in this study is adequate for the purpose.

\section{Analytical estimation of mass loading factor}\label{app:analytical_mload}
In this section, we analytically estimate the mass loading factor for the planar case. For this we choose a plane parallel to galactic disk (as discussed in Section \ref{sec:mload}) at a height of $z$ kpc both above and below the disk. The net mass outflow rate can be considered as the sum of the gas mass flowing out through the shocked CGM (hereafter shell; $\dot{M}_{\rm shell}$) and the shocked wind region ($\dot{M}_{\rm wind}$).
The mass flux carried by the shell can be estimated as
\begin{eqnarray}\label{eq:Mdotshell}
\dot{M}_{\rm shell}& = & 2 \pi r sin\theta \Delta r \rho_{\rm sh} v_{\rm z}\nonumber \\
             & = & 2 \pi \rho_{\rm sh} v_{\rm r} z \Delta r \sqrt{1-\frac{z^2}{r^2}}
\end{eqnarray}
where $r$ is the position of forward shock and $v_{\rm z}$ is the z component of the gas velocity, which can be written as $v_{\rm z} = v_{\rm r} cos\theta = v_{\rm r} \big(\frac{z}{r}\big)$. Here, $\rho_{\rm sh}$ is the gas density of the shell which is taken to be $4$ times the ambient gas density assuming strong shock condition. $\Delta r$ is the width of the shell.

\begin{figure}
\includegraphics[width=3in]{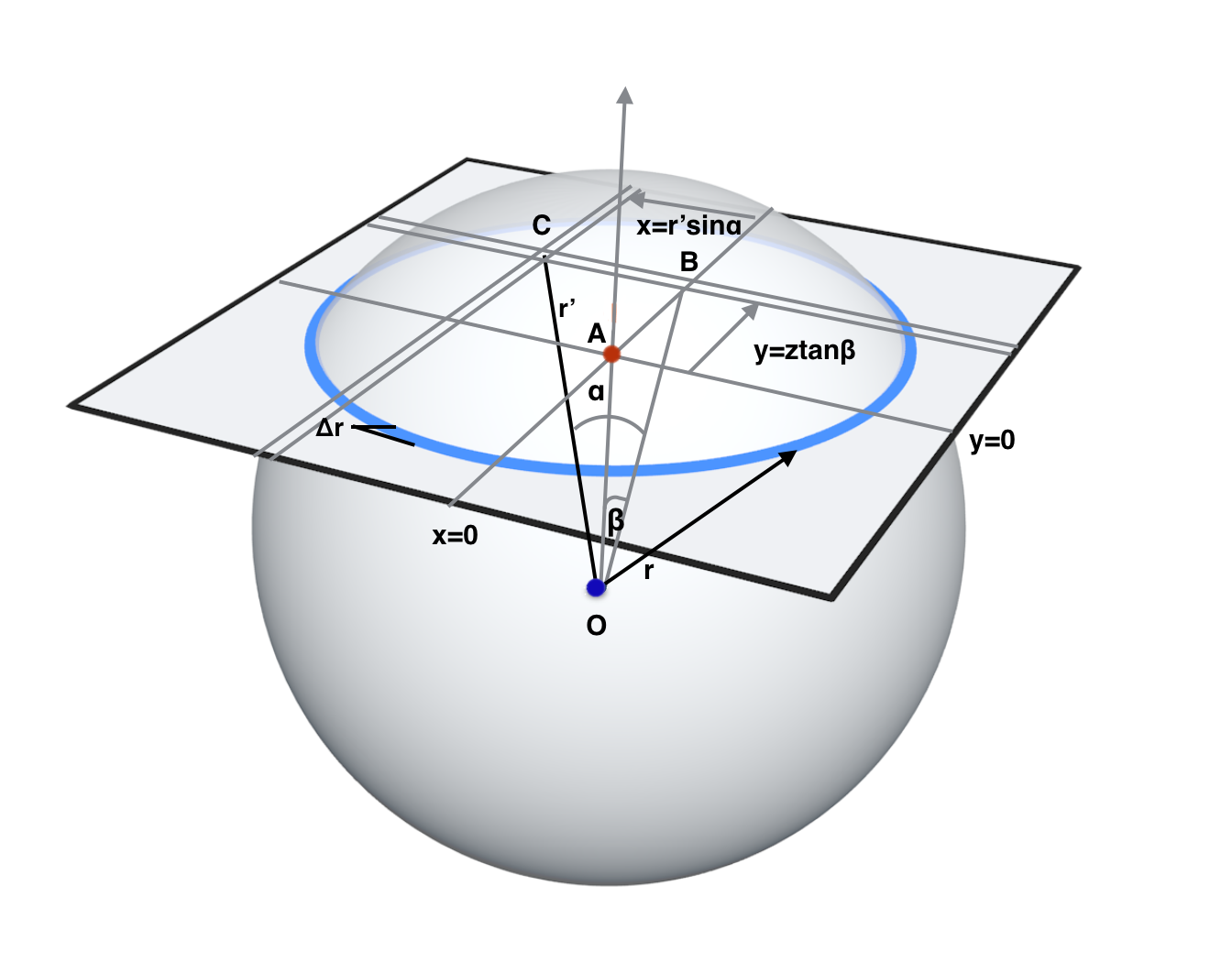}
\caption{Schematic diagram for estimating mass loading factor analytically.}
\label{fig:sch}
\end{figure}

\begin{figure}
\centering
\includegraphics[width=0.5\textwidth]{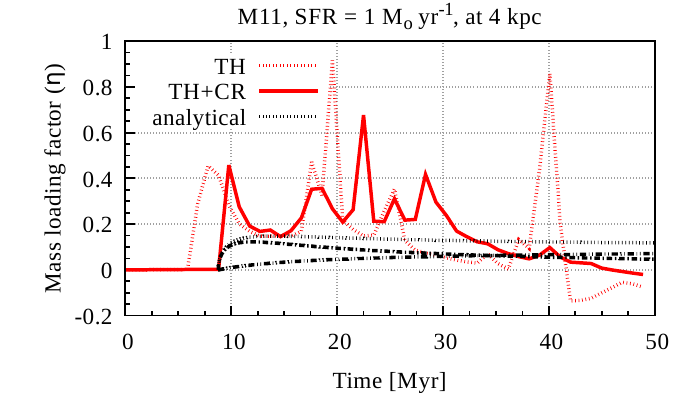}
\caption{Comparison of mass loading factor for M11 galaxy, SFR = 1 $\sfr$ between our simulation (red curves) and analytic method (black curves). The dash-dotted and dash-dot-dotted curves show mass loading by the shocked gas ($\dot{M}_{\rm shell}$; see Eq. \ref{eq:Mdotshell}) and by the wind ($\dot{M}_{\rm wind}$, see Eq. \ref{eq:Mdotwind}) respectively, and the black dotted curve represents the total mass loading factor when the outflow reaches 4 kpc. In the early evolutions, the dash-dot-dotted curve is below the dash-dotted curve because wind is less denser compared to shocked-CGM gas until the shell exceeds $4$ kpc. The red dashed (TH) and solid (TH+CR) curves show the mass loading factor calculated from simulation at a height $z = 4$ kpc. }
\label{fig:comploadana}
\end{figure}

Next, to calculate the mass flux carried by the shocked wind, we define the planar surface to be an integration of differential area elements $dx$ and $dy$. Following the geometry shown in the schematic diagram (Fig. \ref{fig:sch}),  $x=OB \, \tan \alpha={z \over \cos \beta} \tan \alpha$ and $y=z \tan \beta$. We also have ${z \over r'}=\cos \alpha \cos \beta$, since in triangle COB, $\cos \alpha=OB/r^{\prime}$, and in triangle OAB, $OB=(\cos \beta/z)$. Therefore $dx=(z/\cos \beta) \sec^2 \alpha d\alpha$ and $dy=z \sec^2 \beta d\beta$. Also, $v_z=v_r \cos \alpha \cos \beta$. Hence the mass flux carried by the wind is given by 
\begin{eqnarray*}\label{eq:Mdotwind}
\dot{M}_{\rm wind} &=& \int \int \rho_{\rm w}(r^{\prime}) v_{\rm z} dx dy \nonumber\\
&=& 2\times 2 \times {\dot{M}_{\rm inj} \over 4 \pi} \int_{\alpha=0}^{\alpha=\cos ^{-1} {z \over r \cos \beta}} \int_{\beta=0}^{\cos ^{-1}{z  \over r}} \nonumber\\ 
&&   {\cos \alpha \cos \beta \over (r^{\prime})^2} \Bigl ({z \over \cos \beta} sec^2 \alpha \Bigr ) \, \Bigl (z \sec^2 \beta \Bigr )  d\alpha  d\beta \nonumber\\
&=& {\dot{M}_{\rm inj} \over \pi} \int_{\beta=0}^{\cos ^{-1}{z  \over r}} d\beta  \int_{\alpha=0}^{\alpha=\cos ^{-1} {z \over r \cos \beta}} \cos \alpha d\alpha \nonumber\\
&=& {\dot{M}_{\rm inj} \over \pi}  \int \sin \Bigl (\cos ^{-1} {z \over r \cos \beta} \Bigr ) \, d\beta\nonumber\\
&=&{\dot{M}_{\rm inj} \over \pi}\int_{\beta=0}^{\cos ^{-1}{z  \over r}} \sqrt{1-{z^2\over r^2 \cos ^2 \beta}} \,, d \beta \,\qquad r\ge z
\nonumber\\
\end{eqnarray*}
\begin{eqnarray}\label{eq:Mdotwind}
&=&{\dot{M}_{\rm inj} \over \pi}\Bigl [\tan ^{-1}   \Bigl ( {{r \over z} \tan \beta \over \sqrt{{r^2 \over z^2}-1} \sqrt{1-{\tan^2 \beta \over {r^2 \over z^2}-1 }}} \Bigr ) \bigg | 
^{\cos ^{-1} {z \over r}} _0
\nonumber\\
&&\qquad -{z \over r} \sin ^{-1} \Bigl ( {  \tan \beta \over \sqrt{{r^2 \over z^2}-1}} \Bigr ) \bigg | ^{\cos ^{-1} {z \over r}} _0 \Bigr ]\nonumber\\
&=& {\dot{M}_{\rm inj} \over \pi}\Bigl [[{\pi \over 2}-0]  - {z \over r} \sin ^{-1} \Bigl ({ {z \over r} \tan (\cos ^{-1} {z \over r}) \over \sqrt{1-{z^2 \over r^2}}} \Bigr )\Bigr ] \nonumber\\
&=& {\dot{M}_{\rm inj} \over \pi} \Bigl [ [{\pi \over 2}-0]-{z \over r} \sin ^{-1} \Bigl ( { \sqrt{{r^2 \over z^2}-1} \over  \sqrt{{r^2 \over z^2}}-1 }\Bigr ) \Bigr ] \nonumber\\
&=& {\dot{M}_{\rm inj} \over 2}\Bigl (1-{z \over r} \Bigr ) \, 
\end{eqnarray}
\\
A comparison of analytically estimated mass loading factor with our simulation for M11 galaxy (${\rm SFR=1}$ $\sfr$) is shown in Figure \ref{fig:comploadana}. The curves show that the time evolution of mass loading factor in our simulation is qualitatively similar to the above analytic estimates, albeit with some noticeable differences. This is due to various assumptions in our analytic calculation, e.g., the morphology of the outflowing gas (which is assumed as spherical) and shock jump condition. Moreover, the formation of cold clumps may increase the mass loading factor and it is difficult to include all these in analytical calculations self-consistently.

\end{document}